\pdfoutput=1
\documentclass[fleqn,usenatbib,usedcolumn]{mnras}   %a4paper,

\usepackage{txfonts}
\let\la=\lesssim % for less than similar from txfonts, not \la from mnras.cls

\usepackage[T1]{fontenc}
\usepackage{graphicx}	% Including figure files
   \hypersetup{pdfauthor={E. Fern\'{a}ndez-Alvar et al.},
               pdftitle={Chemical trends in the Galactic Halo from APOGEE data},
               pdfkeywords={Galaxy: halo -- Galaxy: stellar content -- stars: abundances},
               bookmarksnumbered=true}
   \volume{{\rm in press}}

\title[Chemical trends in the Halo from APOGEE]{Chemical trends in the Galactic Halo from APOGEE data}

\author[E. Fern\'andez-Alvar et al.]{
E. Fern\'andez-Alvar,$^{1}$\thanks{E-mail: emma@astro.unam.mx}
L. Carigi,$^{1}$
C. Allende Prieto,$^{2,3}$
M. R. Hayden,$^{4}$
T. C. Beers,$^{5}$
\newauthor
J. G. Fern\'andez-Trincado,$^{6}$
A. Meza,$^{7}$
M. Schultheis,$^{4}$
B. X. Santiago,$^{8,9}$
A. B. Queiroz,$^{8,9}$
\newauthor
F. Anders,$^{10,9}$
L. N. da Costa$^{11,9}$   %11,9
C. Chiappini,$^{10,9}$
\\
$^{1}$Instituto de Astronom\'{\i}a, 
              Universidad Nacional Aut\'onoma de M\'exico, M\'exico D.F., M\'exico\\
$^{2}$Instituto de Astrof\'{\i}sica de Canarias,
              V\'{\i}a L\'actea, 38205 La Laguna, Tenerife, Spain\\
$^{3}$Universidad de La Laguna, Departamento de Astrof\'{\i}sica, 
             38206 La Laguna, Tenerife, Spain\\
$^{4}$Laboratoire Lagrange (UMR7293), Universite de Nice Sophia
Antipolis, CNRS, Observatoire de la Cote dAzur,\\
BP 4229, F-06304 Nice Cedex 4, France\\
$^{5}$Department of Physics and JINA Center for the Evolution of the Elements, 
      University of Notre Dame, Notre Dame, IN, 46556, USA\\
$^{6}$Institut Utinam, CNRS UMR 6213, Universit\'e de Franche-Comt\'e,
      OSU THETA Franche-Comt\'e-Bourgogne, Observatoire de Besan\c{c}on,\\
      BP 1615, 25010 Besan\c{c}on Cedex, France\\
$^{7}$Departamento de Ciencias Fisicas, Universidad Andres Bello, Sazie 2212, Santiago, Chile\\
$^{8}$Instituto de F\'isica, Universidade Federal do Rio Grande do Sul, Caixa Postal 15051, 91501-970 Porto Alegre, Brazil\\
$^{9}$Laborat\'orio Interinstitucional de e-Astronomia--LIneA, Rua Gal. Jos\'e Cristino 77, 20921-400 Rio de Janeiro, Brazil\\
$^{10}$Leibniz-Institut fur Astrophysik Potsdam (AIP), An der Sternwarte 16, 14482 Potsdam, Germany\\
$^{11}$Observat\'orio Nacional, Rua Gal. Jos\'e Cristino 77, 20921-400 Rio de Janeiro, Brazil}

\date{Accepted XXX. Received YYY; in original form ZZZ}

\pubyear{2016}

\begin{document}
\label{firstpage}
\pagerange{\pageref{firstpage}--\pageref{lastpage}}
\maketitle

\begin{abstract}

The galaxy formation process in the $\Lambda$-Cold Dark Matter
scenario can be constrained from the analysis of stars in the Milky
Way's halo system. We examine the variation of chemical abundances in distant
halo stars observed by the Apache Point Galactic Evolution Experiment
(APOGEE), as a function of distance from the Galactic center ($r$)
and iron abundance ([M/H]), in the range 5 $\lesssim r \lesssim$
30 kpc and $-2.5 <$ [M/H] $<$ 0.0. We perform a statistical analysis
of the abundance ratios derived by the APOGEE pipeline (ASPCAP) and
distances calculated by several approaches. Our analysis reveals
signatures of a different chemical enrichment between the inner and outer regions of the halo,
with a transition at about 15 kpc. The derived metallicity
distribution function exhibits two peaks, at [M/H] $\sim -1.5$ and
$\sim -2.1$, consistent with previously reported halo metallicity
distributions. We obtain a difference of $\sim 0.1$ dex for
$\alpha$-element-to-iron ratios for stars at $r > 15$ kpc and [M/H] $>
-1.1$ (larger in the case of O, Mg and S) with respect to the nearest halo
stars. This result confirms previous claims for low-$\alpha$ stars
found at larger distances. Chemical differences in elements with
other nucleosynthetic origins (Ni, K, Na, and Al) are also detected. C and N do not provide reliable information about the interstellar medium from which stars formed because our sample comprises RGB and AGB stars and can experience mixing of material to their surfaces.

\end{abstract}


%%%%%%%%%%%%%%%%%%%%%%%%%%%%%%%%%%%%%%%%%%%%%%%%%%

%%%%%%%%%%%%%%%%% BODY OF PAPER %%%%%%%%%%%%%%%%%%

\section{Introduction}
\label{introduction}

The $\Lambda$-Cold Dark Matter paradigm predicts that galaxies form
hierarchically from mergers of lower mass subsystems. Numerical simulations of the formation of Milky Way-like galaxies based on this scenario (e.g., Tissera et al. 2014, and references therein) predict that the halo of our Milky Way is expected to comprise at least two diffuse stellar components with differing
spatial distributions, chemistry, and kinematics, along with a number of individual over-densities and stellar debris streams.  A large body of recent observations of the Milky Way and
external galaxies provide evidence supporting this model. In particular,
the Milky Way's stellar halo has been found to be far from homogeneous
(Belokurov et al. 2009), with a metallicity distribution function (MDF)
that differs between the inner- and outer-halo regions (Carollo et al. 2007,
2010; Beers et al. 2012; Allende Prieto et al. 2014). Chen et al. (2014) and Janesh et al. (2016) have found similar results based on in situ samples of distant giants in the halo. Analyses of relatively local samples of halo stars with photometric metallicity determinations (e.g., An et al. 2013, 2015), combined with available proper motions, have also indicated the presence of significant numbers of stars from the outer-halo population at distances within $\sim$ 10 kpc of the Sun. 

A dichotomy in the $\alpha$-element-to-iron ratios, [$\alpha$/Fe], for
stars with halo kinematics has also been identified in the Solar
Neighbourhood (Fulbright 2002; Gratton et al. 2003; Ishigaki et al.
2010; Nissen $\&$ Schuster 2010, 2011). Since the $\alpha$-elements and
Fe are primarily produced by different stellar progenitors, their
relative abundances can provide constraints on the nature of the
previous generations of stellar populations, such as the initial mass
function (IMF), the star formation rate (SFR), and the efficiency of
star formation in different environments, all of which affect the
production and ejection of these elements to the interstellar medium
(ISM).

In particular, the $\alpha$-elements are synthesized and expelled mainly by
massive stars in the pre-supernova and supernova stages (Type II
supernovae, SNeII), and Fe is largely produced and driven out by binaries
involving low- and intermediate-mass stars during their last stages of evolution
(Type Ia supernova, SNeIa). Different chemical patterns point to stars
born in environments with different IMFs and SFRs. Thus, chemical analysis
of the halo stellar populations can provide information on the Galactic
formation processes.

The advent of large surveys allows us to better characterize the
properties of the stellar populations in the Galaxy. Previous studies
were performed based on samples of a few hundred halo stars in a local
volume. By contrast, current surveys, such as the Sloan Digital Sky
Survey (SDSS; York et al. 2000; Alam et al. 2015), provide data for
hundreds of thousands of stars throughout the halo of the Milky Way.
Specific programs to investigate the Galaxy have been included in SDSS
and its extensions. The most recent sub-survey of this type is the APO
Galactic Evolution Experiment (APOGEE; Majewski et al. 2015). This
program has observed $\sim 150,000$ stars for which stellar
parameters and chemical abundances have been determined. Analysis of
these high-quality data has already confirmed the [$\alpha$/Fe]
dichotomy, exploring nearby halo stars in the metallicity range $-1.2 <$
[Fe/H]\footnote{[X/H]$=\log_{10}(\frac{N(X)}{N(H)})-\log_{10}(\frac{N(X)
}{N(H)})_{\odot}$.} $< -0.55$ (Hawkins et al. 2015). 

The SDSS stellar surveys explore the Galaxy over a broad range of
distances, up to $\sim$100 kpc from the Galactic center. The
aforementioned studies inferred halo properties from stars identified by
their local kinematics; the new data permit investigation of the
properties of the Galactic halo identified by location in the
Galaxy. Analyses of in situ halo stars can provide more complete
information about the halo as a function of distance.

Fern\'andez-Alvar et al. 2015 (hereafter FA15) determined elemental
abundances from low-resolution optical stellar spectra in the SDSS
database, comprising: i) Observations from the original SDSS project and
data from the Sloan Extension for Galactic Understanding and Exploration
program (SEGUE; Yanny et al. 2009) and its extension (SEGUE-2), and ii)
Spectrophotometric calibrators from the Baryon Oscillation Spectroscopic
Survey (BOSS; Dawson et al. 2013). This paper examined the variation of
[Fe/H], [Ca/H], and [Mg/H] as a function of distance from the Galactic
center, $r$, as well as the [Ca/Fe] and [Mg/Fe] abundance ratios as a
function of $r$ and [Fe/H]. Chemical gradients were detected for these
three elements, as well as variations in the [Ca/Fe] and [Mg/Fe] vs.
[Fe/H] behaviours as a function of $r$, pointing to different
$\alpha$-element enrichment histories for the inner- and outer-halo
regions. In this paper, analysis of higher-quality data from APOGEE
enables an independent assessment of these trends based on improved
stellar parameters and chemical abundances. 

This paper is organised as follows. Section~2 provides
a brief description of the APOGEE data. Section~3 describes how we
selected our in-situ halo sample, the stellar parameters and
abundances determined by the APOGEE Stellar Parameters and Chemical
Abundances Pipeline (ASPCAP), the available distance estimates for
APOGEE stars, and the methods used to determine the chemical trends
across the halo system. Section~4 presents our results, which are described in
more detail in Section~5. Finally, we summarise our main conlusions in
Section~6.

\section{Observations}
\label{observations}

Our analysis was performed making use of the DR12 data products for
APOGEE observations taken between September 2011 and July 2014 (Eisenstein et al. 2011; Nidever et al. 2015; Majewski et al. 2015). Using
the same 2.5m telescope at Apache Point Observatory as that employed for
previous SDSS projects (Gunn et al. 2006), APOGEE is a Galactic survey designed to
obtain infrared stellar spectra in the H-band (1.5-1.7 $\mu$m) with a
resolving power of R $\sim 22,500$. From such spectra, stellar
atmospheric parameters and chemical abundances of 15 elements (C, N, O,
Na, Mg, Al, Si, S, K, Ca, Ti, V, Mn, Fe, Ni) were determined with the
ASPCAP pipeline (Garc\'ia P\'erez et al. 2016; Holtzman et al. 2015). APOGEE was designed to
explore the principal stellar components of the Galaxy, mainly the
Galactic disk and bulge, but it also observed stars that are members of
the Galactic halo. Halo stars were targeted following the same general
color-cut criteria, $(J-K)_{0} > 0.5$ as all APOGEE observations. Halo
targets in APOGEE lie mainly at Galactic latitudes b $> 16^{\circ}$. For
further details regarding the target selection in APOGEE, see Zasowski
et al. (2013).

\section{Analysis}
\label{analysis}

The aim of this work is to evaluate the variation of elemental
abundances across the Galactic halo, using in situ halo stars out to
the largest distances reached by the APOGEE observations, $\sim 20-30$
kpc from the Galactic center.

\subsection{Sample}
\label{data}

We first remove stars from our sample with unreliable stellar parameters
and chemical abundance estimates, taking into account the flags provided
in the data files that indicate suspicious ASPCAP results and/or
instrumental issues (see Holtzman et. al 2015). Specifically, we reject
those stars in the database for which the STAR\_BAD bit flag in the
ASPCAPFLAG bitmask is set. This flag warns about stars with unreliable
$T_{\rm eff}$ and $\log g$ estimates, bad matches to synthetic spectra
in the ASPCAP analysis, signal-to-noise ratios per pixel in the final
combined spectrum lower than 50, and/or cases in which the spectrum
exhibits broad lines likely due to significant stellar rotation. In
addition, we do not consider the spectra of stars with the GRIDEDGE\_BAD
flag set in the ELEMFLAG bitmask, which correspond to those stars for
which the resulting abundance estimate is closer than 1/8th of the grid
spacing to the edge of grid (see Garc\'ia P\'erez et al. 2016). Finally, we also avoid stars with spectra
affected by persistence in the detectors (which may lead to significant
errors in stellar parameters and abundance determination -- see Section
5.7 in Holtzman et al. 2015), by considering only stars that do not have
the PERSIST\_LOW, PERSIST\_MED and PERSIST\_HIGH flags set in the
STARFLAG bitmask. For more details about APOGEE flags we refer the
interested reader to the web page
http://www.sdss.org/dr12/algorithms/bitmasks/.

Besides the selection criteria discussed above, we also apply other
restrictions in $T_{\rm eff}$ and $\log g$ that can arise from issues
described by Holtmann et al. (2015). We only consider stars with estimated
$T_{\rm eff}$ > 4000~K, because at cooler temperatures the quality of
the ASPCAP fitting is significantly lower. The calibration performed to
the $\log g$ FERRE outputs, by comparing with asteroseismic $\log g$
estimates for stars observed by APOGEE in the Kepler field (Pinsonneault
et al. 2014), shows that stars at $\log g \geq 4$ deviate considerably
from asteroseismic gravities (Holtzmann et al.). Therefore, they only
calibrated data with lower $\log g$ estimates. Thus, we only consider
stars with surface gravity estimates in the range $1.0 < \log g < 3.5$.
In addition, we reject stars that were targeted as belonging to open or
globular clusters, since we are interested in the chemical analysis of
halo field stars; stars in clusters can exhibit chemical patterns that
differ from those observed in field stars (see, e.g., Lind et al.
2015; Fern\'andez-Trincado et al. 2016). 

Finally, in addition to the existing target selection criteria for
APOGEE observations, we select our halo sample by considering objects
with derived distances from the Galactic plane $|z| > 5$ kpc. The
resulting sample comprises a total of $\sim 400$ stars.

In order to check whether our sample comprises only stars belonging to
the halo, we also inspect their kinematics. For this purpose, we derive the
full space velocities respect to the local standard of rest, $V_{\rm
tot}$, using the radial velocities provided by DR12 and proper motions
from
UCAC4\footnote{http://www.usno.navy.mil/USNO/astrometry/optical-IR-prod/ucac}
(Zacharias et al. 2013). Stars with $V_{\rm tot}$ $>$ 180 $\rm km
s^{-1}$ are usually considered to belong to the halo. Our sample
includes some stars with lower $V_{\rm tot}$. It is not clear why some
of these stars have such low velocity values. One possibility is that, at a few kiloparsecs from the Sun, the UCAC4 proper motions uncertainties are similar or greater than the intrinsic proper motions (see Section 6.3 in Bovy et al. 2014). These uncertainties propagates to the derived velocities, introducing large errors. After having checked that excluding these stars
does not significantly impact our results, we have decided to
retain them in our sample. 

The top panel in Figure~1 shows the MDF for our final sample of stars,
which is discussed in Section 4.1.
We note that our MDF is in agreement with previous MDFs derived
for halo samples (Carollo et al. 2007,2010; Allende Prieto et al. 2014; Chen et al. 2014;
An et al. 2013, 2015),
displaying a maximum at [M/H] $\sim -1.5$ and a secondary
peak at lower [M/H] ($\sim -2.1$). We conclude that our
sample is comprised almost entirely of bona-fide halo stars. 

\begin{figure}
\center
\includegraphics[scale=0.5]{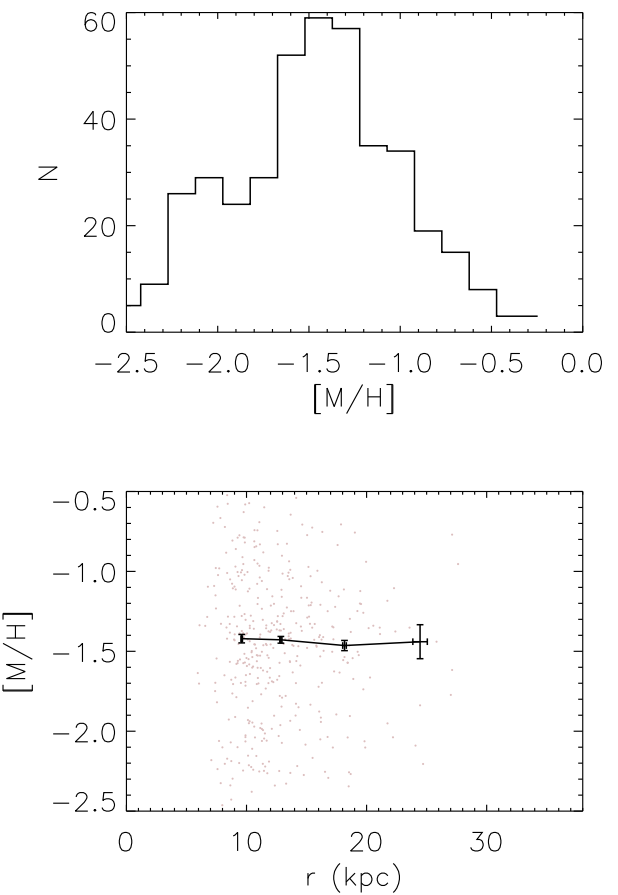}
\caption{Top panel: Metallicity distribution function derived from the calibrated [M/H] for our sample of 410 halo stars with $|z| > 5$ kpc. Bottom panel: Median [M/H] (calibrated) as a function of the distance form the Galactic center, $r$, calculated with distances by the Brazilian Participation Group --see Section \ref{distances}-- (from the peak of their second PDF) for the same sample.}
\label{mdf}
\end{figure}

\subsection{Stellar parameters and chemical abundances}
\label{abundances}
 
The basic techniques followed in ASPCAP for stellar parameter and chemical abundance determination
are the same as in the analysis performed by FA15 -- comparison of the observed spectrum with a
library of synthetic spectra covering a range of stellar parameters,
looking for the parameter combination that returns the lowest $\chi^2$.
This comparison is performed using the code FERRE\footnote{FERRE is
available from http://hebe.as.utexas.edu/ferre} (Allende Prieto et al.
2006). The analysis proceeds in two steps:

\begin{enumerate}

\item The stellar parameters $T_{\rm eff}$ and $\log g$ are determined
from a search fitting the entire available spectral range, and

\item Individual chemical abundances are derived by searching only in the [Fe/H] dimension,
with the $T_{\rm eff}$ and $\log g$ fixed at the previously determined
valuoes, and fitting isolated spectral windows dominated by features of the element of interest. 

\end{enumerate}

ASPCAP includes several improvements, and performs a more refined
abundance determination than FA15. For instance, the synthetic grid
includes separate [C/Fe], [N/Fe], and [$\alpha$/Fe] dimensions, and the
atmospheric models in the synthetic spectra generation are consistent
with the variations in C and the $\alpha$-elements abundances. An
improved atomic line list is used, and other upgrades (broadening to
account for macroturbulent velocity, etc.) are considered (for more
details see Garc\'ia P\'erez et al. 2016). Most importantly, the higher
S/N ($> 100$) and resolving power (R $\sim 22,500$) of APOGEE spectra
allow for an improvement of the accuracy of estimates compared with
those obtained from the lower-resolution optical spectra. The spectral
features resolved in the near-infrared H-band also permit the
measurement of many more chemical elements. On the other hand, APOGEE
was designed to observe mainly the Galactic disk and bulge. For this
reason, the survey targeted very few halo stars at distances farther
than 30 kpc from the Sun. Therefore, we cannot explore the trends in the
most distant regions of the halo investigated in FA15, which
included stars with Galactocentric distances beyond 40 kpc.

$H-$band stellar spectra generally exhibit weaker lines than optical spectra. With a minimum opacity at the transition between the dominance of continuum H$^{-}$ bound-free and free-free opacity at about 1.6 $\mu$m, in the center of the $H-$band, photons escape from deeper atmospheric layers in the $H-$band than in the optical spectral of late-type stars. Deeper layers are warmer and produce weaker absorption lines, and $H-$band transitions tend to have higher excitation, which makes them weaker as well. Fewer and weaker lines, even though they are less dependent on the choice of micro-turbulence, means more limited information in the spectra. In addition, metal-poor atmospheres have higher gas pressure, increasing the role of line damping, and a reduced opacity enhances departures from Local Thermodynamical Equilibrium. These effects may limit the accuracy and precision of the APOGEE abundances for metal-poor stars more than for their solar-metallicity couterparts.

As in FA15, we would like to evaluate how the individual elemental
abundances vary with distance from the Galactic center and stellar
metallicity. In FA15, we took our individual iron abundance measurements
([Fe/H]) as an indicator of the metallicity, [M/H], in the stars. In the
present paper we also consider this elemental abundance as the primary
estimate of stellar metallicity.

The variation of
the iron abundance with respect to the Solar value is considered in ASPCAP as a
dimension of the synthetic library. All the other elements, except C, N,
and the $\alpha$-elements, change in the same proportion as iron with
respect to Solar abundances. ASPCAP provides two estimates for the iron abundance. On the one hand, an iron abundance measurement
([M/H]) is obtained from the fit of the entire available APOGEE spectral
range, which includes spectral features from several chemical elements.
On the other hand, another estimate ([Fe/H]) is derived by seeking
the best match in the [M/H] dimension, but fitting only spectral windows
containing iron lines (Garc\'ia P\'erez et al.
2016). Both measurements are expected to be quite close to one another. 

A systematic over-estimate at low metallicities was detected in Holtzman et al. 2015 for both
[M/H] and [Fe/H] by comparing with [Fe/H] measurements from the
literature. Consequently, they performed an external calibration to [M/H] (a
second-order fit) that corrects for this effect, but this was not
applied to [Fe/H]. Moreover, each individual element was internally calibrated
independently from the others to remove abundance trends with effective
temperature in open clusters.

In the case of C, N, and the $\alpha$-elements, ASPCAP calculates their
variation over Fe by directly searching within the library. We use these
quantities when evaluating [X/Fe] for these elements. For the other
elements, we calculate [X/Fe] ratios from internally-calibrated individual chemical abundances
(including [Fe/H]). It is not yet clear what might be
the cause of the [Fe/H] systematic deviation at low metallicities, and
other individual abundances may be affected as well. However, ratios in
the form [X/Fe] from measurements with the same systematic deviation
cancels this effect.

We are interested in evaluating differences in the behaviours of individual
elemental abundances. The [M/H] determination is influenced by the contribution
of elements other than iron, which can induce deviation from the true
iron abundance. Consequently, the [X/M] ratios may not be reliable
for our purposes, so we avoid their use in this paper. 

Additionally, we are
interested in evaluating the chemical trends in different metallicity
bins. We choose the calibrated [M/H] as our indicator of the global
metallicity, because it is corrected for the over-estimation on the
metal-poor side. [Fe/H] is unsuitable in this case, because it is still
affected by the systematic deviation. Considering it to derive trends
with metallicity would place metal-poor stars in higher metallicity
bins, and the resulting trends would be distorted. Thus, we use
internally-calibrated abundance estimates when discussing abundance ratios, but
employ the externally-calibrated [M/H] to set our metallicity scale.

Notice that the analysis in Holtzman et al. (2015) revealed hints of "some issue that may be affecting the reliability of the ASPCAP [Ti/H] abundance"; and a large scatter in [Na/H] and [V/H], which also lead to be aware of the limited precision of these abundance estimates. For these reasons, we cautiously interpret the resulting trends for these elements.

\subsection{Distances}
\label{distances}

A number of independent groups have been working on the derivation of
distance estimates for APOGEE stars, which we consider in our
present analysis; these are described in Santiago et al. (2016), Hayden et al.
(2015), and Schultheis et al. (2014). 

The derivation of distances for APOGEE giant stars necessarily involves dealing with stars with a very wide range of luminosities, increasing the susceptibility to uncertainties in the stellar evolution models adopted. Nevertheless, comparison across different implementations and with Gaia/Hipparcos parallaxes suggest that no significant systematic errors are present in the distances adopted in this paper.

Distances derived by the SDSS-III Brazilian Participation Group (BPG; Santiago et al. 2016) were computed using the Bayesian methodology explained in Burnett \& Binney (2010), Burnett et al. (2011), and Binney et al. (2014). From the measured spectroscopic parameters coupled with 2MASS photometry, they obtained the posterior distance probablility distribution function (PDF) for each star over a grid of PARSEC (Bressan 2012) stellar evolutionary models. Their model
prior includes information such as the spatial distribution of stars in our Galaxy and the initial mass function.

The BPG considered the ASPCAP [M/H] calibrated values, except for
metallicities [M/H] $> 0.0$, because in this regime stars may be
``over''-calibrated, due to the choice of a second-order fit to the data
values running away at the edges of its range of validity. They also
applied an additional surface gravity calibration with respect to the
DR12 $\log g$ values for stars belonging to the red clump. This does not
affect our sample because we only consider stars with $\log g$ < 3.5,
which are not members of the red clump. The accuracy of their results
was tested with simulations and previous distance estimates for several
samples of observations from the literature. The statistical distance uncertainties are at a level of 20\%. Although we cannot completely exclude this possibility, there are no strong indications of systematic distance biases towards large distances (low gravities).

Hayden et al. (2015; H15) derived distances following the same
methodology as the BPG. They compared the stellar parameters from ASPCAP
with PARSEC isochrones from the Padova-Trieste group (Bressan et al.
2012), considering matches within 3$\sigma$. They then computed the
probability distribution function of all distance moduli in the range
between the minimum and maximum magnitudes matching the isochrone grid.
As for the BPG estimates, the precisions are at a
level of 15-20\%.

Finally, the methodology followed by Schultheis et al. (2014; S14)
consisted of comparing with Padova isochrones from Marigo et al. (2008),
and looking for those that match most closely with the ASPCAP parameters $T_{\rm
eff}$, $\log g$, and calibrated [M/H]. These authors recognized that
not taking into account the $\alpha$-element enhancements
and the use of Solar-scaled isochrones may introduce errors in their
distance estimates. Thus, they estimated the median precisions in their derived
distances to be on the order of $\sim30-40$\%. 

From these three sets of distance estimates, we determine distances from
the Galactic center, $r$, as follows:

\begin{equation}
r = \sqrt{d^{2} + R_{\odot}^{2} - 2d R_{\odot} \cos{b} \cos{l}}
\end{equation}

\noindent and the distance from the Galactic plane, $z$,

\begin{equation}
z = d\sin{b},
\end{equation}

\noindent where $b$ and $l$ are the Galactic coordinates, 
provided in the APOGEE data files, and $R_{\odot}= 8.0$ kpc, given by Ghez et al.
(2008).

\subsection{Evaluation of the chemical trends}
\label{evaluation}
%We carried out two different and complementary evaluations. 

We now consider the variation of individual chemical abundances, as a function
of distance from the Galactic center, for each of the 15 elements
determined by ASPCAP. Our sample covers the range 5 $\lesssim r
\lesssim$ 30 kpc. We inferred their trends by calculating the median
[X/H] in bins of $\Delta r =5$ kpc or wider, assuring a minimum of 100
stars per bin. Figures~1 (bottom panel) and 2 show the resulting
trends. 

We are also interested in examining how the elemental abundance to iron ratios vary with
$r$ and [M/H]. For this purpose, we split our sample into three
metallicity bins ($-2.5 <$ [M/H] $<-1.8$, $-1.8 <$ [M/H] $< -1.1$, and
$-1.1 <$ [M/H] $< 0.0$), and calculate the median ratios for stars at
$r < 10$, $10 < r < 15$, and $r > 15$ kpc, in each one of the three
metallicity ranges. The choice of these bins satisfies our aim to
calculate the median ratios from the largest possible data sets, in
order to infer the chemical trends as accurately as possible. Figures~3 and 4 show the resulting median [X/Fe] ratios, as a function of $r$ and
[M/H], respectively, evaluated separately in the corresponding
metallicity and distance bins. 

We indicate with error bars the median absolute deviation (MAD) divided by the square root of the number of points from which we derive each median abundance (we assume that the uncertainties follow a Gaussian distribution). The abundance dispersion known for the halo is $\sim 0.5$ dex (Allende Prieto et al. 2014). The bulk of the [X/H] and [X/Fe] uncertainties are $\sim 0.1$ with few exceptions ([Na/H] and [V/H]), and in no cases exceed 0.3 dex, on average, in each bin. Consequently, our sample should be dominated by the natural halo abundance dispersion. However, we also estimate the weighted mean with the uncertainties provided in the APOGEE database, in order to test that the resulting trends are not significantly distorted due to the abundance errors.  

In order to quantify the variation,
Table~\ref{var BPG2p} shows the difference between each median [X/Fe]
ratio with the nearest stars median [X/Fe], $r < 10$ kpc, for each range of
[M/H] considered and with the lowest metallicity median, $-2.5 <$ [M/H] $< -1.8$, for each range of $r$. When the difference is significant (as demonstrated by
application of a Kolmogorov-Smirnov test -- see Section
\ref{verifications}), it is indicated with red text.

Finally, we verify whether the resulting trends are consistent when
taking into account the three sets of available distance estimates. For this
purpose, we analyse whether the variance of the median [X/Fe] ratios,
calculated from the distance estimates by the several groups
described in Section~\ref{verifications}, follows the same trends
inferred from an individual set of estimates.

\section{Results}
\label{results}

\subsection{[X/H] vs. $r$}

We evaluate the median [Fe/H] values, as a function of $r$, using distances
calculated by BPG, H15, and S14. All produce fairly similar results, but we choose to employ the BPG estimates derived from the peak of the second probability distribution function in their analysis (BPG2p - see Santiago et al. 2016). The reason is that it has the least amount of scatter in the [M/H] vs. $r$ relation, and showed little signs of a gradient, which is in agreement with that observed by FA15.

Considering the [M/H] values externally calibrated with [Fe/H] abundances from the literature, the resulting
median values (bottom panel in Figure~1) are around
$\sim -1.5$, which is consistent with the previous works. As mentioned above in Section \ref{data}, we have derived the MDF for
our sample from the calibrated [M/H], shown in the top panel of Figure~1.
The peak of the distribution is
around [M/H] $\sim -1.5$. In addition, a second peak around
$-2.1$ is observed. This is very close to the median metallicity value
associated with the outer-halo region (Carollo et al. 2010; Allende
Prieto et al. 2014; FA15).

The trend of the median [X/H] ratios with distance from the Galactic center,
shown in Figure~2, exhibit constant or decreasing trends. Those inferred for the $\alpha$-elements are fairly constant, except [Mg/H] and [Ti/H], which show a significant decrease, $\sim$ 0.1 dex, from $r < 10$ to $r > 15$ kpc. [C/H] also exhibits a significant variation, the largest among all the elements evaluated, decreasing by $\sim$ 0.2 dex from $r$ $< 10$ to $10 < r < 15$ kpc. 

It is important to recall, however, that the abundances of [C/H] and [N/H] can be altered due to
mixing events in cool red giants, the dominant spectral class of our
sample. Thus, these elements are not reliable indicators of the ISM
chemistry from which these stars formed. Although we provide their resulting median values and trends, we will not comment on them because we are interested in those abundances that provide information of the previous stellar populations.   

The [Mn/H] abundance exhibit a decreasing trend, contrary to the other iron-peak [Ni/H], which does not show significant variation. The elements [Na/H] and [Al/H] both exhibit decreasing trends, although [Na/H] has a larger variance with distance ($\sim 0.4$ dex). [Al/H] decreases $\sim 0.1$ dex
from $r < 10$ kpc to $10 < r < 15$ kpc. Both elements are produced by massive stars and low- and intermediate-mass stars (LIMS); however the production and ejection efficiencies for each element are different. Finally, the [K/H] and [V/H] both
show constant trends.

\begin{figure*}
\center
\includegraphics[scale=0.58,angle=180]{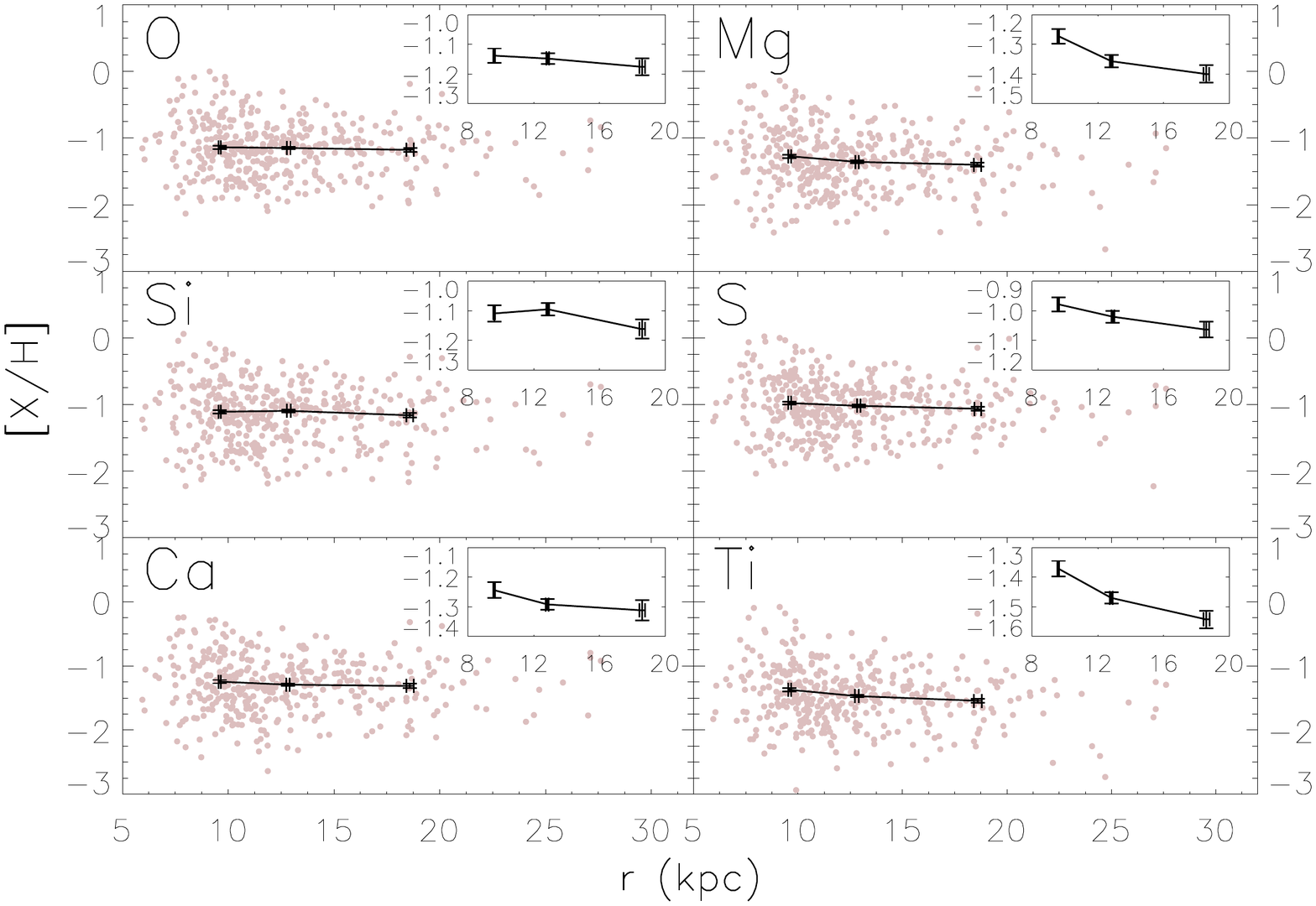}
\includegraphics[scale=0.58,angle=180]{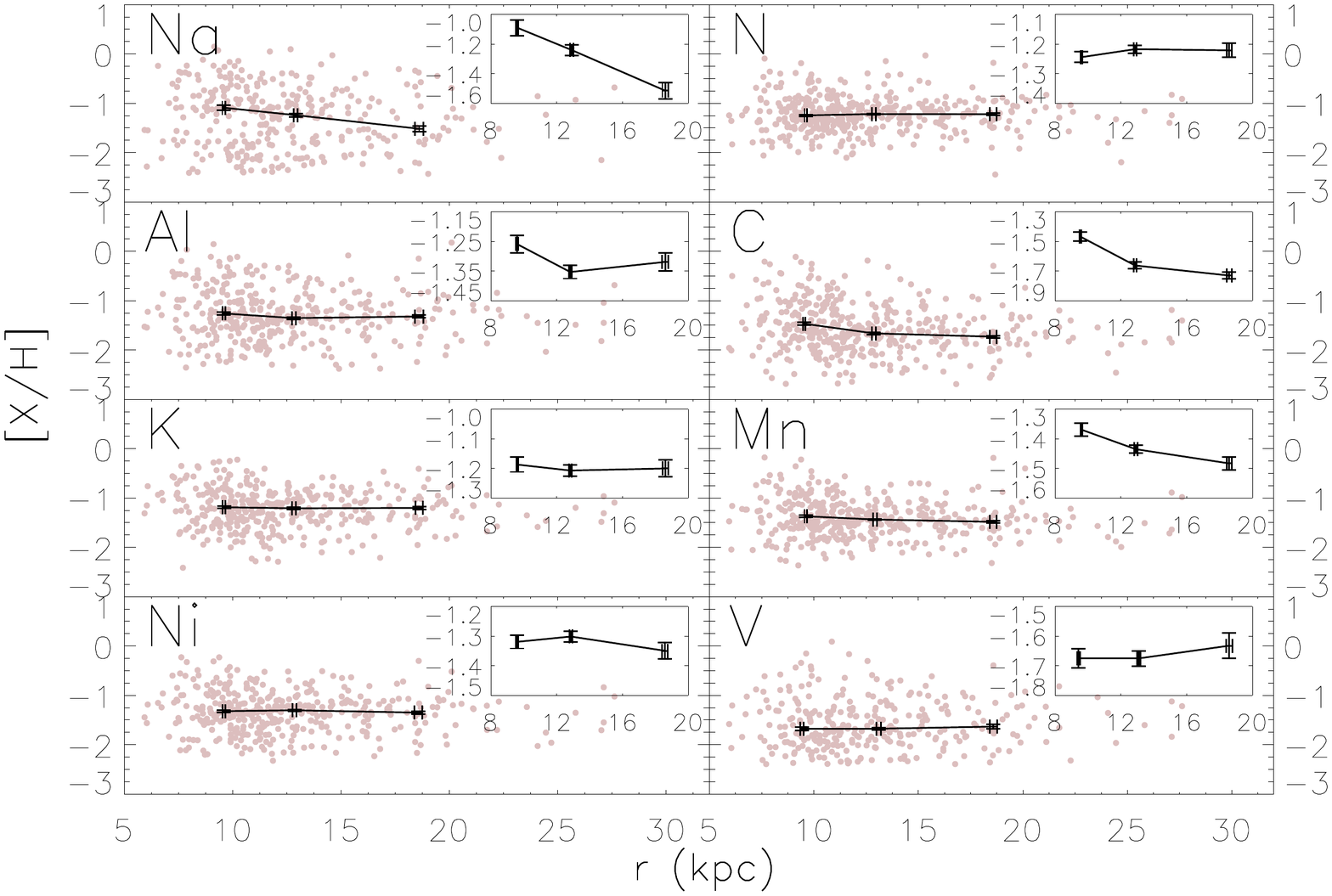} 
\caption{[X/H] median values as a function of the distance from the Galactic center, $r$. 
The top panels correspond to the $\alpha$-elements and the bottom panels to the other analysed elements. The median values with axes covering narrower ranges that emphasize their behaviour are shown as insets at the upper right in each panel.}
\label{xhr}
\end{figure*}

We also evaluate [X/H] trends with $r$, splitting the sample in bins of
[M/H]. As expected, [X/H] is higher as [M/H] increases. Overall, the
elemental abundances exhibit similar variations with $r$ and [M/H]. The
most metal-rich stars exhibit the largest variation with $r$, with
higher median values for stars in the inner-halo region compared to
those in outer-halo region, excepting [Si/H] and [Ti/H] among the
$\alpha$ elements, and [N/H], [Ni/H] and [V/H], which
remain constant.

\subsection{[X/Fe] vs. $r$}

Inspection of the variation of [X/Fe] with $r$ reveals that the chemical trends
depend on the metallicity range considered, as seen in
Figure~3. For the $\alpha$-elements, metal-poor stars have
enhanced [X/Fe] across all $r$. The median ratios decrease as the
metallicity increases. The most metal-rich stars show the
largest variation with $r$, decreasing farther from the
Galactic center. This decreasing trend tends to flatten toward lower
[M/H], although [Ca/Fe] shows a significant fluctuation of 0.07 dex with $r$ at $-2.5 <$ [M/H] $< -1.8$. [Ti/Fe] decreases 0.14 dex in this metallicity bin.

At $-1.1 <$ [M/H] $< 0.0$, [O/Fe], [Mg/Fe], and
[S/Fe] exhibit the largest variation, $\sim 0.2$ dex, with $r$. The [Si/Fe], [Ca/Fe] and [Ti/Fe] ratios decrease by almost $0.1$ dex between the inner and outer
regions. As
was found by FA15, [Ca/Fe] exhibits a larger dependence with metallicity
than [Mg/Fe]. The decreasing trends of [Ca/Fe] with $r$ in each of the
metallicity bins analysed are consistent with the FA15 results, although
with an offset in the median values. On the contrary, the increasing
trends observed in FA15 for [Mg/Fe] at [M/H] $> -1.1$ are not confirmed in this work,
where we find that the median ratio decreases.

The iron-peak elements Ni and Mn do not exhibit ratios that strongly vary
with $r$. Metal-poor stars show enhanced [Mn/Fe], the largest of $\sim 0.13$ dex at $10 < r < 15$ kpc. Noticeably, [Ni/Fe] tends to
decrease ($\sim 0.08$ dex) with distance for stars in the most metal-rich bin. This
pattern is the same observed for the $\alpha$-elements, in agreement
with previous findings (Nissen $\&$ Schuster 2010, 2011; Yamada et al.
2013; Hawkins et al. 2015), although they detected the pattern from the
analysis of [X/Fe] as a function of metallicity.

[Na/Fe] and [Al/Fe]
do not follow the same trends with $r$ and [M/H]. The [Na/Fe] ratio
decreases with $r$, and exhibits
higher median values as metallicity decreases. On the contrary, [Al/Fe]
exhibits a different pattern depending on the distance bin and
metallicity range considered. In the inner regions, $r < 10$ kpc, [Al/Fe] is higher as
metallicity increases. This ratio decreases with $r$ for the most
metal-rich stars, while it tends to increase for the most metal-poor
stars. As a consequence, stars in the outer region have [Al/Fe] that
does not depend so significantly on metallicity than for stars in the
inner region. The theoretical Na and Al yields predicts similar [X/Fe] vs. [Fe/H] behaviours. However, observations in the Solar Neighborhood do not completely follow the theoretical predictions (C\^{o}t\'{e} et al. 2016). Our analysis also reveals a disagreement in [Na/Fe] and [Al/Fe] trends with $r$ and [M/H].

The ratio [K/Fe] also tends to decrease with $r$ at [M/H] $\sim -1.1$, but
stars at [M/H] $< -1.1$ exhibit constant trends. The difference in
[K/Fe] with [M/H] is higher for stars in the outer region. [V/Fe]
increases with $r$ in the most metal-poor stars, and tends to flatten as [M/H] increases, although no well-defined trends with metallicity are observed. The chemical analysis of V should be taken with caution,
however, because its measurement is less reliable (V is determined
exclusively from very weak spectral features -- see Holtzmann et al.
2015.)

\begin{figure*}
\center
\includegraphics[scale=0.58,angle=180]{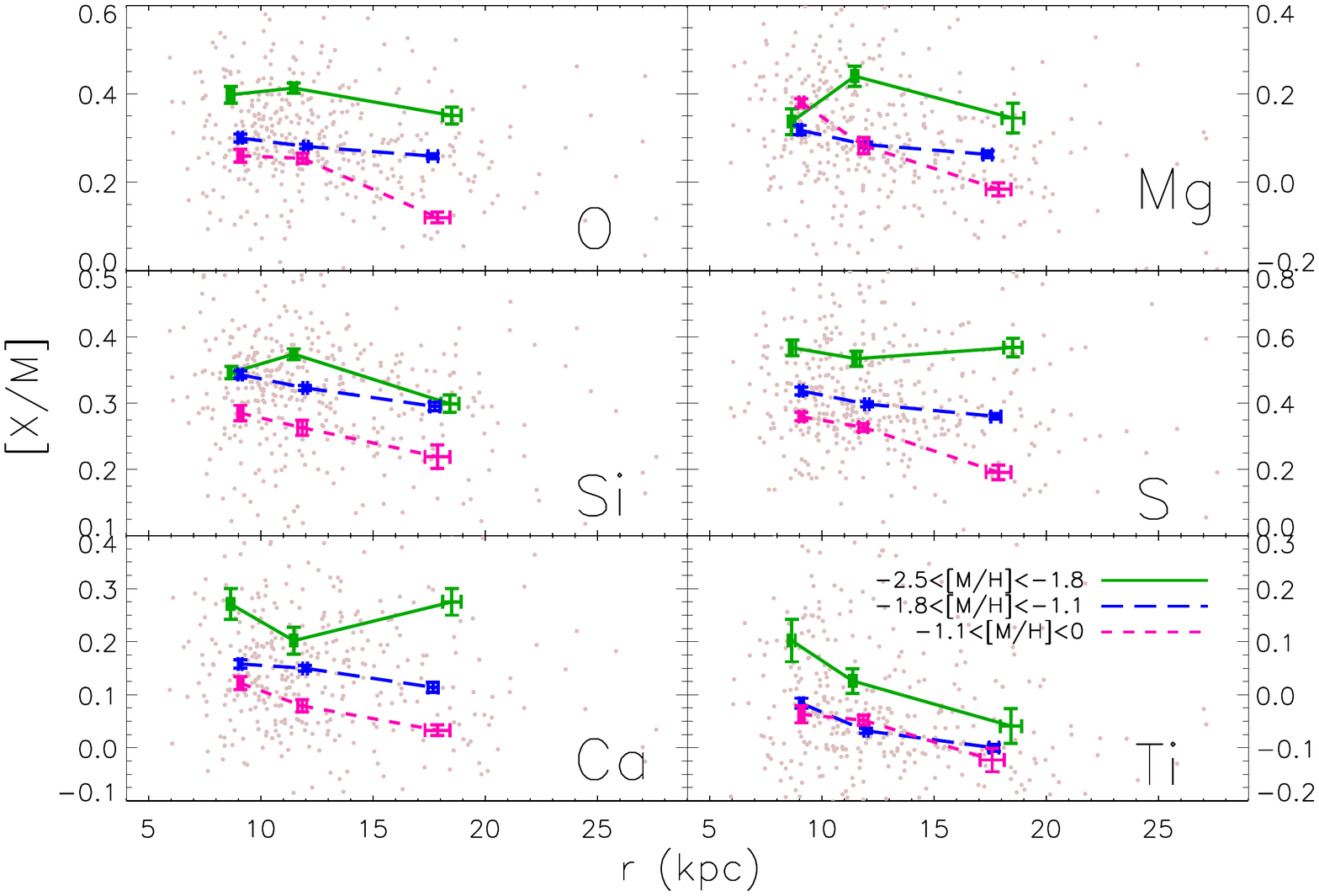}
\includegraphics[scale=0.58,angle=180]{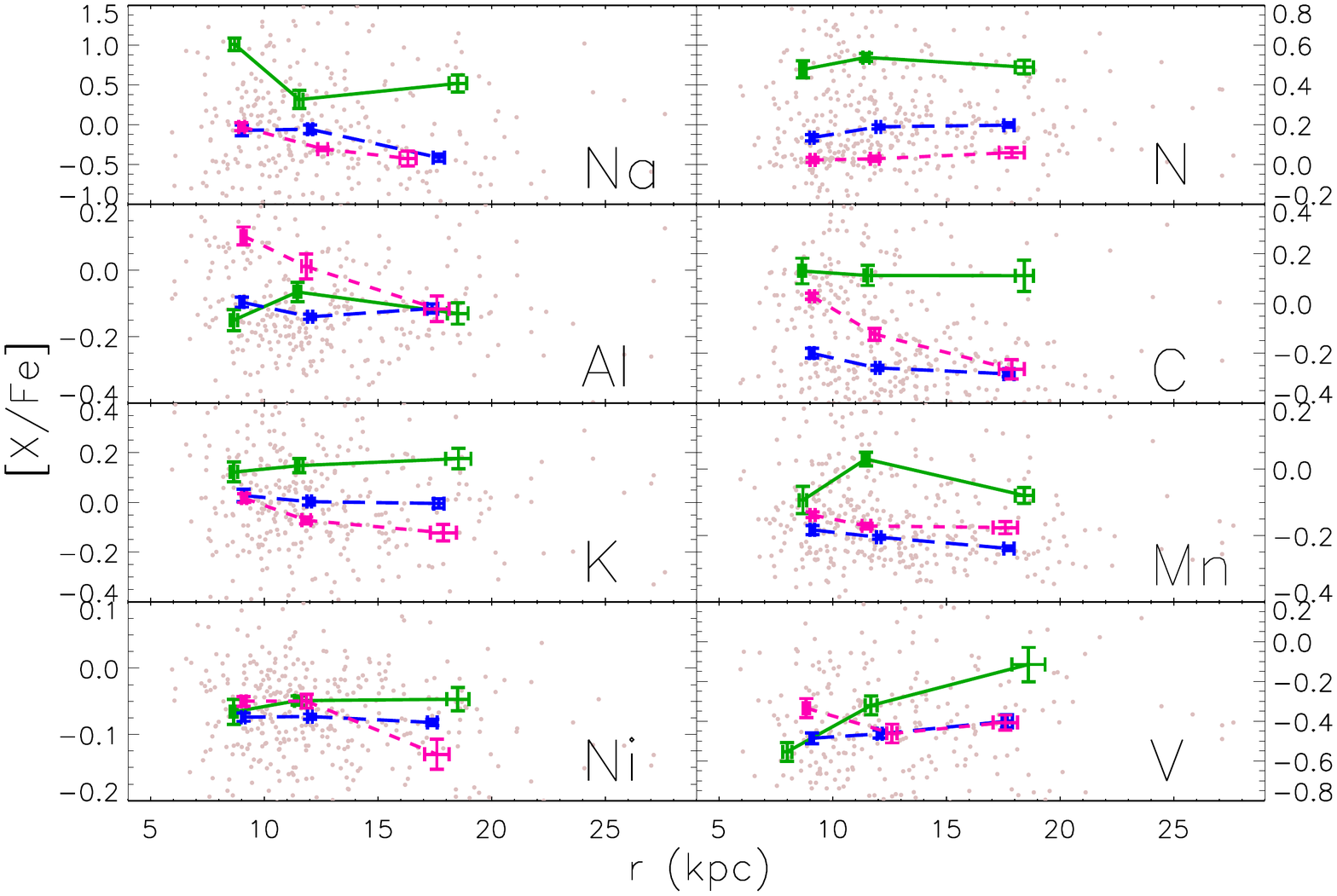} 
\caption{[X/Fe] median values as a function of $r$. The trends are
inferred by splitting the sample into the three [M/H] bins shown in the
legend of the lower right panel in the top set of plots.}
\label{xfer}
\end{figure*}

\subsection{[X/Fe] vs. [M/H]}
%\subsubsection{$\alpha$ elements}

The resulting curves of the median [X/Fe] values, calculated as a
function of [M/H], are shown in Figure~4. Overall, $\alpha$-elements show decreasing trends with [M/H], with few exceptions. Stars at $r < 10$ kpc exhibit a decrease larger than 0.1 dex in [X/Fe] toward higher [M/H], except for [Mg/Fe] and [Si/Fe]. These abundances remain constant and slightly vary at [M/H] $\sim 1.1$, increasing 0.04 dex and decreasing 0.06 dex respectively.   

As $r$ increases, the trends become steeper (except for [Ti/Fe], which tends to flatten). The most distant stars show decreasing variations $> 0.1$ dex for [Mg/Fe], $> 0.2$ dex for [O/Fe] and [Ca/Fe], and $> 0.3$ dex for [S/Fe]. [Si/Fe] and [Ti/Fe] also decrease with [M/H], although less ($< 0.1$ dex). Figure 4 clearly shows the spread in [$\alpha$/M] for stars at [M/H] $> -1.1$ as a function of distance from the Galactic center described in the previous section. This spread, $\geq 0.1$ dex, is similar to the differences observed by Nissen \& Schuster (2010) for [$\alpha$/Fe] as a function of [Fe/H].

Based on the Kolmogorov-Smirnov test, no significant variations larger than 0.1 dex are detected in [Ni/Fe] with [M/H]. However, Figure 2 reveals lower [Ni/Fe] as $r$ increases at [M/H] $\sim -1.1$. The [Mn/Fe] ratio decreases with [M/H] from the most metal-poor stars up
to [M/H] $\sim -1.5$, and increases slightly toward higher
metallicities. The increasing trend on the higher metallicity side is
independent of distance; all the stars in the sample have similar median
[Mn/Fe] ratios. In contrast, the metal-poor tail suggests enhanced
[Mn/Fe] ratios for more distant stars.

There is an overall decrease of [Na/Fe] with [M/H]. The [Na/Fe] ratio 
exhibits the largest variation, but also a large scatter, likely due to the
difficulty of measuring Na from the APOGEE spectra (Holtzmann et al.
2015). The variation of [Al/Fe] with [M/H] clearly depends on the distance bin
considered. The nearest stars exhibit an increasing trend, with
variations of $\sim 0.25$ dex between stars with [M/H] $< -1.8$ and [M/H] $>
-1.1$. This trend tends to flatten with $r$. The [Al/Fe] ratio for stars
at $r
> 15$ kpc is nearly constant. Thus,
for [M/H] $> -1.1$, the median [Al/Fe] decreases with $r$. The metal-poor
stars suggest an opposite trend with $r$.

The median [K/Fe] ratios also reflect a different enrichment pattern
that depends on Galactocentric distance. Trends are steeper as $r$ increases, with
enhanced ratios in the metal-poor tail and lower values toward the
highest metallicities that we consider. At [M/H] $\lesssim -1.5$, there is
no significant difference in the median ratios calculated for the three distance
bins. 

We find an increasing trend in [V/Fe] with distance
for stars at $r < 10$ kpc, flattening as $r$ increases. Our resulting ratios have lower median [V/Fe] ratios for
all the stars in our sample. Overall, distant stars have higher [V/Fe],
and the trends for the three $r$ bins merge for stars with [M/H] $>
-1.1$. However, as mentioned above, estimates of the V abundance are
less reliable due to its weak features in APOGEE spectra.

\begin{figure*}
\center
\includegraphics[scale=0.58,angle=180]{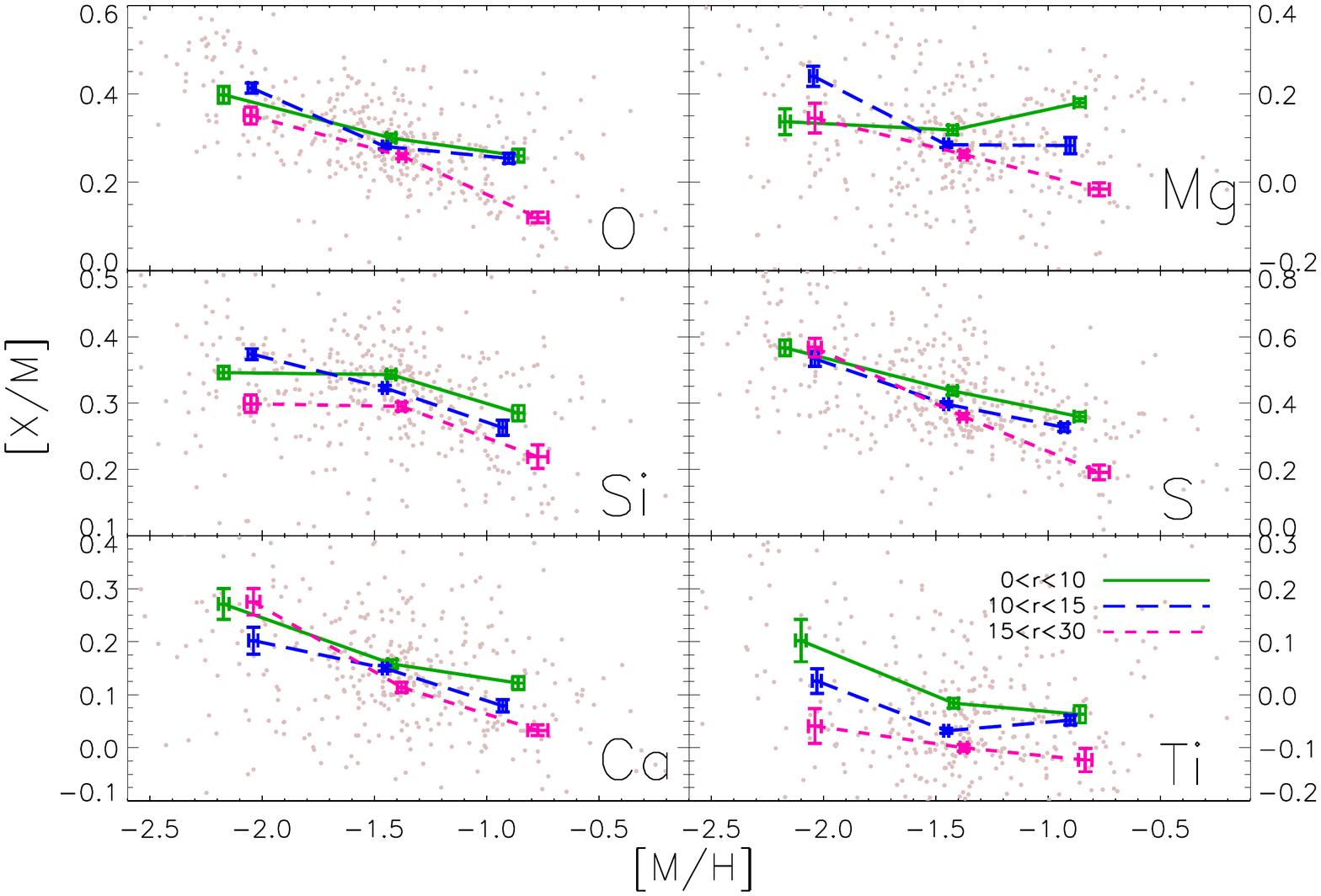}
\includegraphics[scale=0.58,angle=180]{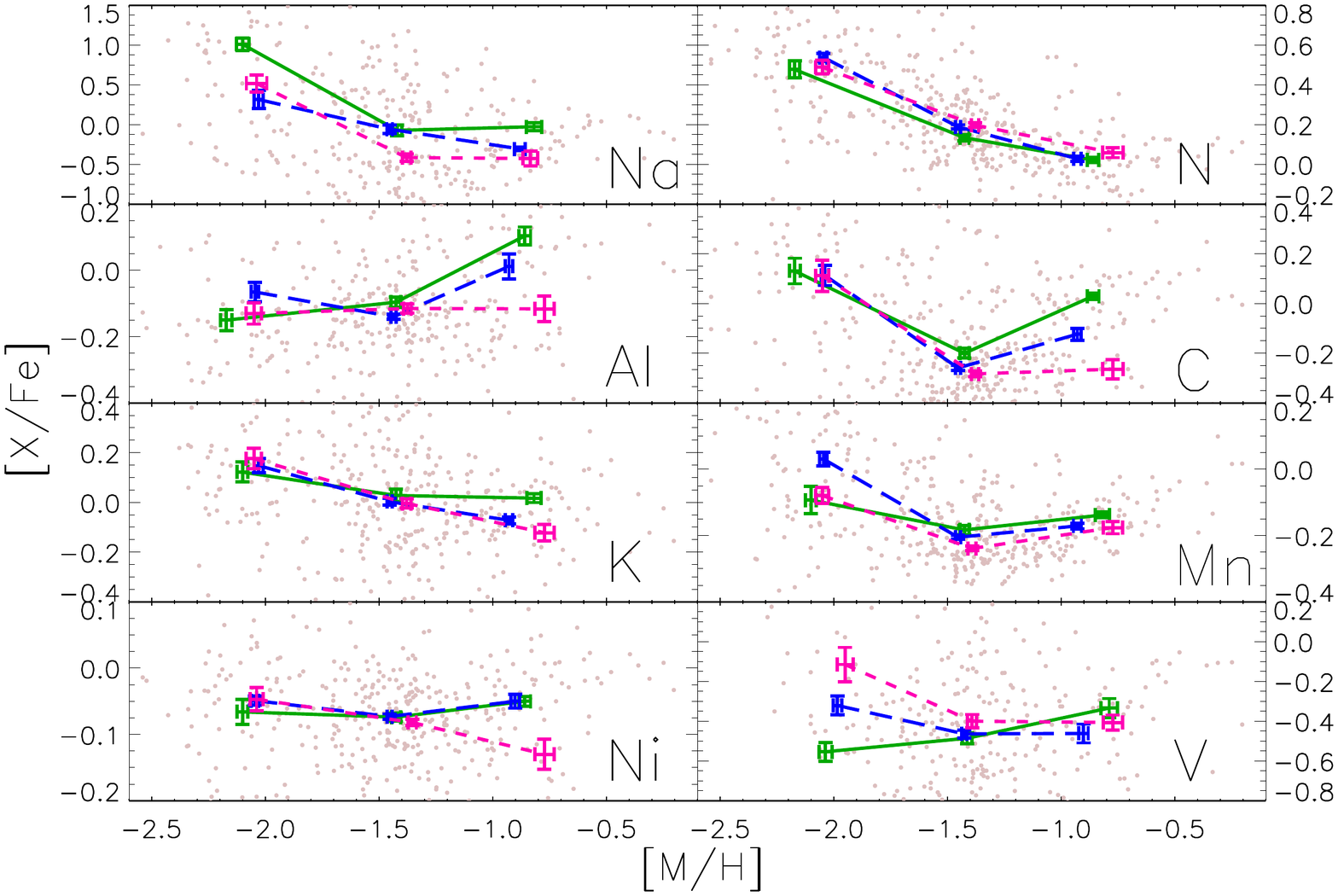} 
\caption{[X/Fe] median values as a function of [M/H], splitting the
sample into three $r$ bins: $r<10$ kpc, $10<r<15$ kpc, and $r>15$ kpc.}
\label{xfemh}
\end{figure*}

\subsection{Validation of trends}
\label{verifications}

In order to check whether the resulting trends reported above depend on
the chosen distance estimates, we calculate the median abundance ratios
for the six sets of distances available for DR12 APOGEE data. We
first calculate the median and its variance for each of the six median abundance
ratios in the corresponding $r$ and [Fe/H] bins, and then evaluate how these
variances change with $r$ and [M/H]. Overall, these curves confirm the
previous inferred trends. Thus, the different distance estimates lead to
the same qualitative trends, although the particular median values
differ slightly depending on the set of distances considered.

As an additional check, we carried out the previous evaluations by considering the mean ratios weighted with the measurement errors -- the resulting trends are qualitatively similar
to the median ratio curves. We also performed a
Kolmogorov-Smirnov test in order to verify if the observed differences
between our median ratios over bins in $r$ and [M/H] are statistically
significant. We proceed by calculating the cumulative distribution
function (CDF) for each bin in $r$ and [M/H] evaluated, and the maximum
difference between each CDF and the CDF corresponding to the lowest $r$
bin over each [M/H] range (and to the lowest [M/H] bin over each $r$ range), then compare with the critical values of the
K-S statistic. In order to consider a difference significant, we demand
that it cannot be rejected at higher than the 10\% level. We perform the test to evaluate variations of [X/H] vs. $r$ and [X/Fe] vs. $r$ and [M/H]. The last four columns in Table \ref{var BPG2p} shows the resulting variations in [X/Fe] as a function of $r$ and [M/H] and the level of significance obtained for them; those with a significant difference are indicated in red. In the previous sections, we have only described significant variations after applying this test. 

Notice that we derive [X/Fe] (except in the case of C, N and $\alpha$-elements) from [Fe/H] measurements, which exhibits a systematic deviation comparing with literature values. We do not know whether [X/H] is affected by the same systematic deviation. If this is the case, the deviation would be absent in the resulting [X/Fe]. On the contrary, the ratios would be systematically underestimated toward lower metallicities. In most of the cases the observed [X/Fe] trends with metallicity are the opposite. The slopes would be higher if [X/Fe] were underestimated.

Uncertainties in the stellar parameters could lead to systematic errors in the chemical abundances and thus distortions in the inferred chemical trends. M\'esz\'aros et al. 2015 investigated about the deviations in [Fe/H], [C/Fe], [N/Fe], and [$\alpha$/Fe] (each of the ASPCAP $\alpha$-elements) due to uncertainties in $T_{\rm eff}$. From their Figure 3 we observe that the most sensible ratio to a $T_{\rm eff}$ variation is [O/Fe]. A $\Delta T_{\rm eff} \sim 200$ K would imply a $\Delta \log g \sim 0.6$ dex in the RGB, and an uncertainty in [O/Fe] $\sim$ 0.25 dex (lower for the other abundances of our interest). The peak of the $\log g$ distribution in the $r < 10$ and $r > 15$ kpc bins shift toward 0.5 dex lower, aproximately. Considering the number of the stars in our furthest bin, we derive that the possible systematic error due to uncertainties in the parameters would lead to underestimate 0.08 dex our [O/Fe]. However, we detect a larger variation at the most metal-rich side.

The APOGEE sample comprises stars in the red giant branch (RGB) and possibly in the asymptotic giant branch (AGB) stages of evolution. These stages are reached by LIMS, producing, at the same time, heavy elements with diverse efficiencies, depending on the initial mass and metallicity of the stars. The photospheres of these stars are enriched mainly by carbon, nitrogen, fluorine, and heavier elements synthetized by the slow neutron capture process (the $s$-process) and by  proton-capture nucleosynthesis (the $p$-process). The mixing from the interior (core and shells surrounding the core) to the stellar envelope results in self-pollution of the stellar photosphere.

The abundances of all the $alpha$-elements analised in this paper (O, Mg, Si, S Ca, Ti)
are representative of those abundances in the interstellar medium from which the stars formed,
because these chemical elements are not synthetized and are not carried to the photosphere of LIMS,
(see review by Karakas \& Lattanzio 2014). Moreover, half of the other elements studied in this work (K, V, Mn, Ni ) are also not generated by  LIMS. Therefore, our results for O, Mg, Si, S, Ca, Ti, K, V, Mn, Ni are independent of the evolutionary status of the APOGEE stars.
 
Nevertheless, C, N, Na, and Al are produced by LIMS, but in very different proportions. Among these four elements:
\begin{enumerate}
\item C is the most produced, mainly by low-mass stars ($M < 3.5 M\odot$) during the thermal pulses and the third dredge-up (during the AGB).
\item N is the second most produced, mainly by intermediate-mass stars ($M > 3.5$) during the first and second dredge-ups (during the ascent of the RGB and AGB, respectively) and hot bottom burning.
\item Na and Al are synthetized mainly by stars with $M > 3.5$ during the second and third dredge-ups, with Na more abundantly produced than Al.
\end{enumerate}

Since C and N are mainly produced by LIMS, we caution that C and N do not reliably represent of the original stellar abundances. The Na and Al abundances might be enhanced at high Galactocentric radii, because:
\begin{enumerate}
\item These elements are produced by intermediate-mass stars in evolved stages (AGB) and consequently by more luminous stars, and
\item The APOGEE sample at large $r$ may be more weighted towards these more luminous objects.
\end{enumerate}

However, the $r$-trends of Na and Al (Figures 2 and 3) appear to show no enhacenment at large radii.

Stars at different distances could also have different age distributions: at the bottom of the giant branch stars may be biased toward a different age distribution than at the upper side, which might affect the overall chemical trends. Nearer stars would be biased toward slightly younger and more metal-rich stars and further stars toward older and more metal-poor ones: however, if it were the case, further stars would show higher [$\alpha$/Fe] ratios, because older stars would have formed from an ISM mainly enriched by SNII. Our analysis reveals the opposite trend with galactocentric distance.

Finally, we estimate the impact of the distance errors in our sample. We assume a normal distribution with uncertainties of $20\%$, and add this noise to the BPG distance values. The resulting fraction of stars with $|z_{n}| > 5$ (after adding the noise) but $|z| < 5$ (without noise), respect to the stars with $|z| > 5$, is $40\%$ for $r < 10$ kpc, $13\%$ for $10 < r < 15$ kpc, and $3\%$ for $r > 15$ kpc. However, these fractions reduce to $1\%$ and lower if we consider stars at $|z_{n}| < 4$. This means that there could be a contamination of $\sim 40\%$ of stars at $4 < |z| < 5$ kpc in our $r < 10$ bin. At this $|z|$ range, the density of thin disc stars is negligible, of thick disc stars is $\sim 2\%$ the density of stars in the plane, and a little less for the density of halo stars. Thus, $50 -70\%$ of the contaminant stars, i.e., a $\sim 20-30\%$ of the stars in the $r < 10$ kpc bin, are likely to belong to the thick disc. The resultant median abundances would be dominated by halo stars. Besides, previous works have not found differences in chemistry between thick disc and halo stars. Therefore, we assert that the chemical trends would not be greatly distorted by this contamination.

\section{Discussion}

Differences for a number of the chemical trends with [M/H] are clear 
between stars at $r < 10$ and $r > 15$ kpc. The lower
$\alpha$-element-to-iron ratios found at larger distances are consistent
with the low-$\alpha$ population reported during the past decade by a
number of workers (e.g., Fulbright 2002; Gratton et al. 2003; Nissen \&
Schuster 2010, 2011; Ishigaki et al. 2012, 2013; Hawkins et al. 2015).
From the kinematical properties of their samples, they estimated that
the orbits of these stars would place them farther away than stars
exhibiting higher [$\alpha$/Fe] ratios. This is consistent with our
finding (and that of FA15).

Our present study finds a decrease with $r$
consistent with that obtained by FA15 at $r < 20$ kpc. They also
observed a considerably larger drop occurring at a Galactocenctric
radius between $20 < r < 40$ kpc, which we cannot confirm in this work,
as the APOGEE observations do not extend to cover this distance range.

The trend observed with [M/H] for the low-[$\alpha$/Fe] population has been
interpreted in terms of Type Ia SNe, which contribute iron but little
$\alpha$-elements. The differences in [X/Fe] observed in the present work between stars at $r < 10$
kpc and $r > 15$ kpc for the $\alpha$-elements ($\sim 0.1$ dex for
Si, Ca, and Ti, and higher for O, Mg, and S) are consistent with their
expected relative contributions in SNeIa explosions (Tsujimoto et al.
1995) -- higher for Si ($17\%$) and Ca ($25\%$) than for Mg,
O, and S (negligible).

It is envisioned that this population would form later,
during a long period of a relatively slow star formation, from
an ISM polluted by both Type II SNe and Type Ia SNe. The
high-[$\alpha$/Fe] population would form much earlier, during a short
period of rapid star formation, as these stars originated from an ISM
enriched mainly by Type II SNe. If the star-formation history presents
several bursts, the high-$\alpha$ stars would form during the early
bursts, while the low-$\alpha$ stars would form at the beginning of
later bursts from a ISM contaminated by recent SNeIa, as appears to have
occurred in dwarf spheroidal galaxies (see, e.g., Carigi et al. 2002).
Nissen \& Schuster (2010) also proposed that the low-$\alpha$ population
could be born in systems that were later accreted into the Milky Way's
halo, and which had experienced a long star-formation history.

However, the decrease in [$\alpha$/Fe] ratios due to the contribution of
SNeIa encounters difficulty in explaining the pattern of some abundance
ratios, in particular, the decrease observed for [Ni/Fe] with
metallicity for the low-$\alpha$ population, and the absence of
different [Mn/Fe] ratios between both populations. These two elements are expected to be released by SNeIa; thus,
different patterns should be detected in Ni and Mn with respect to iron
between stars formed in an ISM enriched by SNeIa and stars that formed
from gas without their contribution. In fact, it would be expected that
low-$\alpha$ stars would have higher [Ni/Fe] and [Mn/Fe] than the
high-$\alpha$ stars. However, this is not seen either in our work or in previous studies.

A possible explanation was suggested by Kobayashi et al. (2006), who claimed that, for an IMF biased toward stars that explode as low-mass SNIe-II, this would lead to lower [$\alpha$/Fe]. Kobayashi et al. (2014) also proposed that the nucleosynthesis of
10-20 M$_{\odot}$ stars could explain the difference in the [$\alpha$/Fe]
ratios detected in halo stars. Interestingly, McWilliam et al. (2013) also claimed that a
`top-light' IMF might provide an explanation for the [$\alpha$/Fe] and [Eu/Fe]
deficiencies found in the analysis of the M54 cluster belonging to the
Sagittarius dwarf spheroidal galaxy. This possibility would also explain the [Mn/Fe] and
[Ni/Fe] patterns, but it would imply a complex IMF behaviour to explain
the different trends -- a metallicity dependent IMF for the low-$\alpha$
stars and a metallicity independent IMF for the high-$\alpha$
population, as Nissen \& Schuster (2011) indicated.

Another remaining issue is the fact that the nucleosynthetic
contribution from AGB stars (the immediate
progenitors of the white dwarfs involved in Type Ia SNe explosions)
should be detected in the low-$\alpha$ population, if SNeIa have
contributed their iron. However, these stars exhibit {\it lower} [Na/Fe]
ratios than the stars presumed to be formed prior to the contributions
from SNeIa. As Nissen et al. (2014) speculated, this could be the result if the progenitors of the low-$\alpha$ population
were stars of intermediate mass (4-8 M$_{\odot}$), which contribute
little C to the ISM, and even less Na and Al.
The lower median [Na/Fe] and [Al/Fe] ratios that we found for stars at
$r > 15$ kpc are consistent with a slow chemical-enrichment history. No
signature of significant enrichment from AGBs (neither Na nor Al) are
detected in the distance bin where the low-$\alpha$ population
dominates.

The observed increase in [Al/Fe] observed for stars in the
inner regions can be explained with the assumption of
metallicity-dependent yields for massive stars (Nomoto et al. 2013), and
the flat trend with [M/H] for the farthest stars by the cancellation of
the increase in Al by the even higher contribution of Fe from SNeIa.
However, we do not detect an increasing trend of [Na/Fe] with [M/H] in
the nearest stars, as observed by Nissen \& Schuster (2010) for the
high-$\alpha$ population, which is well-explained by the same
metallicity-dependent yields also invoked for Na (Nomoto et al. 2013).
Observationally, Na is determined by APOGEE from relatively weak
lines in the H-band, yielding less accurate measurements, as suggested
by the large scatter detected in our sample as well as in Holtzman et
al. (2015), when inspecting [Na/Fe] vs. [Fe/H]. For these reasons, our Na results
should be considered with caution.

The higher [X/Fe] ratios at metallicities lower than $-1.8$ but lower ratios at higher metallicities, may be explained by the
combination of a top-heavy IMF and a slower star-formation rate (SFR) in the subsystems
accreted by the Galaxy. Recent cosmological simulations predict that
massive satellites merging with the host galaxy contribute at smaller
radii than low-mass systems (Amorisco 2015). Low-mass systems are
expected to experience outflows that release their gas. This would
prevent, or at least greatly suppress, subsequent star formation. The expected IMF for low-mass stellar systems at
low metallicities is characterized by discontinuities, i.e., a lack of
some massive stars (Cervi\~{n}o 2013). The signature of the few high-mass
stars in such low-metallicity environments produces stochastic effects on
the abundance ratios (Carigi \& Hern\'andez 2008), which could explain the
higher dispersion observed in [X/Fe] for metal-poor stars.

The stars that we observe today at metallicities lower than $-1.8$, born in these low-mass subsystems, would be formed from
the nucleosynthetic contribution from a few very massive stars, leading to high ratios with respect to iron at very low
metallicities, followed by less-massive stars slowly contributing to the
ISM where the current stars were born. On the contrary, the inner
regions would be formed in an ISM that would have reached the same
metallicity faster, with the contribution of a larger number of massive
stars, although the upper mass limit of the IMF would be lower (Yamada et al. 2013).

\section{Conclusions}
\label{conclusions}

We have analysed a sample of $\sim 400$ halo stars targeted by the
APOGEE survey, located at $|z| > 5$ kpc from the Galactic plane, and evaluated the chemical
trends for the 15 individual abundances determined by ASPCAP. In order
to be sure that our trends were not unduly influenced by the estimated
distances to our stars, we made use of the available distances estimated
by three independent methods for APOGEE stars. Our main conclusions are
the following:

\begin{enumerate}

\item An analysis of the elemental abundance ([X/H]) variation 
with distance from the Galactic center, $r$, up to the farthest
distances observed by APOGEE ($\sim 20-30$ kpc), revealed that the
chemical trends are almost constant or decrease with $r$. The variation mainly occurred
for stars with a global metallicity [M/H] $> -1.1$. 

\item We confirmed that the qualitative chemical trends inferred from
our data do not depend on the considered distance set.

\item The resulting iron abundance trend calculated from the calibrated [M/H] parameter is constant across the range of $r$ examined, 
$5 \lesssim r \lesssim 30$ kpc. The variation for nearer stars
measured in our analysis is barely lower than that observed in the
previous analysis of in situ halo stars performed by FA15 with SDSS
optical spectra at lower resolution. They also reported a larger decrease taking
place at $20 < r < 40$ kpc. Our evaluation of [Fe/H] and [Ca/H] are consistent with their results,
but we cannot probe the chemical trends at $20 < r < 40$ kpc due to a
lack of sample stars in this distance range.

\item The median calibrated [M/H] values,
$\sim -1.5$, also agrees with previous reports for inner-halo stars
(Carollo et al. 2007, 2010; Chen et al. 2014; FA15). The derived MDF
from [M/H] also shows a second peak at [M/H] $\sim -2.1$,
which resembles the median metallicity value associated with the outer-halo population (Carollo et al. 2007, 2010; Beers et al. 2012).

\item We also evaluated the trends in the abundance ratios relative to
iron, [X/Fe], with distance from the Galactic center from in situ stars for 14 chemical
elements. This analysis shows that a population of stars with different
$\alpha$-element enrichment becomes dominant beyond $r \sim 15$ kpc:

a) For the $\alpha$-elements, we found significantly lower ratios for more
distant stars at metallicities [M/H] $> -1.1$. We observed a larger
separation in these two populations for [O/Fe], [Mg/Fe] and [S/Fe], but all the
$\alpha$-elements show a significant decrease of $\gtrsim 0.1$ dex.  

Our [Ca/Fe] results with $r$ and [M/H] are consistent with the results reported
by FA15 based on SDSS optical spectra. On the contrary, we found a
decreasing trend of [Mg/Fe] with distance, which disagrees with what they
observed. 

Our results are also consistent with the two different halo populations
reported in APOGEE data by Hawkins et al. 2015 at $-1.2 <$ [M/H] $<
-0.55$. The [O/Fe] and [Mg/Fe] trends we oberved are in agreement with their
work and other previous reports. Conversely, we found that the two
populations also exhibit different [S/Fe] ratios. 

b) We find hints of low-$\alpha$ stars having lower [Ni/Fe]. We detect
different [Mn/Fe] ratios between the inner and outer regions for the most
metal-poor stars ([M/H] $< -1.8$), although a larger sample of stars at
distances farther than 15 kpc and more accurate measurements are
necessary to confirm this result.

c) The [K/Fe] trends with $r$ and [M/H] also provide evidence for different
chemical abundance patterns in stars at $r < 10$ kpc and $r > 15$ kpc.

d) Both [Na/Fe] and [Al/Fe] reveal different chemical patterns for the
nearer stars compared with the more distant stars. The [Na/Fe] ratio
exhibits different
trends with $r$ and [M/H] than found for [Al/Fe], but the Na measurements are
less reliable. The [Al/Fe] ratio increases with metallicity for
inner-halo stars, while the more distant stars exhibit a flat trend.

\end{enumerate}

This work corroborates the suggestion that stars with low [$\alpha$/Fe]
ratios are predominant at larger distances than stars with higher
[$\alpha$/Fe] ratios, in agreement with previous work that inferred the
distances for the low- and high-$\alpha$ populations based on their
kinematical properties (Nissen \& Schuster 2010, 2011; Ishigaki et al.
2010). The lower [$\alpha$/Fe] ratios are consistent with iron
enrichment due to SNeIa; the [Al/Fe] chemical patterns are also
consistent with this hypothesis. The [Ni/Fe] and [Mn/Fe] exhibit trends
with metallicity in both populations that are also consistent with
these previous studies. However, their chemical patterns, as well as the
lack of signatures of AGB enrichment, are not those expected in a
scenario where SNeIa had time to explode. The characteristics of the
environments where both populations were formed remains unclear.

In conclusion, the chemical trends inferred for stars ranging over
distances from the Galactic center of $5 < r < 30$ kpc suggest that, at
$r > 15$ kpc, a stellar population begins to dominate which formed with
a different chemical-enrichment history than stars at $r <10$ kpc.
Characterization of the different stellar populations with a larger
sample of stars will better constrain the IMF and SFR associated with
these previous stellar populations. High-quality data for stars at $r$
farther than 15 kpc will help to clarify the chemical properties of the
more distant stellar populations in the Galactic halo. Alternatively, the
identification of nearby halo stars that probe to large distances (on
the basis of their extreme kinematics) will also permit an increase in the number of
suitable outer-halo stars for further analysis.

\section*{acknowledgements}

E.F.A. acknowledges support from DGAPA-UNAM postdoctoral fellowships.
L.C. thanks for the financial supports provided by CONACyT of M\'exico
(grant 241732), by PAPIIT of M\'exico (IG100115, IA100815) and
by MINECO of Spain (AYA2010-16717). T.C.B. acknowledges partial support
for this work from grants PHY 08-22648; Physics Frontier Center/Joint
Institute for Nuclear Astrophysics (JINA), and PHY-1430152; Physics
Frontier Center/JINA Center for the Evolution of the Elements
(JINA-CEE), awarded by the US National Science Foundation.
Funding for SDSS-III has been provided by the Alfred P. Sloan
Foundation, the Participating Institutions, the National Science
Foundation, and the U.S. Department of Energy Office of Science. The
SDSS-III web site is http://www.sdss3.org/.

SDSS-III is managed by the Astrophysical Research Consortium for the
Participating Institutions of the SDSS-III Collaboration including the
University of Arizona, the Brazilian Participation Group, Brookhaven
National Laboratory, University of Cambridge, Carnegie Mellon
University, University of Florida, the French Participation Group, the
German Participation Group, Harvard University, the Instituto de
Astrofisica de Canarias, the Michigan State/Notre Dame/JINA
Participation Group, Johns Hopkins University, Lawrence Berkeley
National Laboratory, Max Planck Institute for Astrophysics, Max Planck
Institute for Extraterrestrial Physics, New Mexico State University, New
York University, Ohio State University, Pennsylvania State University,
University of Portsmouth, Princeton University, the Spanish
Participation Group, University of Tokyo, University of Utah, Vanderbilt
University, University of Virginia, University of Washington, and Yale
University.

%%%%%%%%%%%%%%%%%%%%%%%%%%%%%%%%%%%%%%%%%%%%%%%%%%



%\bibliographystyle{mnras}
%\bibliography{bibliography} % if your bibtex file is called example.bib

%\begin{thebibliography}{99}
%\end{thebibliography}

%%%%%%%%%%%%%%%%%%%%%%%%%%%%%%%%%%%%%%%%%%%%%%%%%%

%%%%%%%%%%%%%%%%% APPENDICES %%%%%%%%%%%%%%%%%%%%%

\appendix

%\section{}

\begin{table*}
\caption{Median [X/Fe] and $r$ with their corresponding median absolute
deviation (MAD), evaluated in the three [M/H] and $r$ bins, and the
difference between each median and that corresponding to the lowest $r$
bin over each [M/H] range and to the lowest [M/H] bin over each $r$ range.
The significant differences indicated by the Kolmogorov-Smirnov test are in red, followed by the level of significance ($los$).}
\label{var BPG2p}
\begin{tabular}{cccccccccccccc}
\hline
elem & $\rm[M/H]_{l}$ & $\rm[M/H]_{u}$ & $r_{\rm l}$ & $r_{\rm u}$ & N & $<\rm[X/Fe]>$ & $e_{<\rm[X/Fe]>}$ & $<r>$ & $e_{<r>}$ & $\Delta <\rm[X/Fe]>_{[M/H]}$ & $los$ (\%) & $\Delta <\rm[X/Fe]>_{r}$ & $los$ (\%) \\
\hline
\hline
  O &  -2.50 &  -1.80 &   5 &  10 &  21 &    0.40 &    0.02 &    8.65 &    0.15  &  - & - & - & - \\
    &  -2.50 &  -1.80 &  10 &  15 &  47 &    0.41 &    0.01 &   11.46 &    0.14 &    \textcolor{red}{0.01} & 1 & - & - \\
    &  -2.50 &  -1.80 &  15 &  30 &  22 &    0.35 &    0.02 &   18.51 &    0.42 &   \textcolor{red}{-0.05} & 5 & - & - \\
    &  -1.80 &  -1.10 &   5 &  10 &  52 &    0.30 &    0.01 &    9.07 &    0.09  &  - & - & \textcolor{red}{-0.10} & 1 \\
    &  -1.80 &  -1.10 &  10 &  15 & 108 &    0.28 &    0.00 &   12.01 &    0.10 &  -0.02 & 15 & \textcolor{red}{-0.13} & 1 \\
    &  -1.80 &  -1.10 &  15 & 30 &  60 &    0.26 &    0.01 &   17.67 &    0.22 & \textcolor{red}{  -0.04} & 1 & \textcolor{red}{-0.09} & 1 \\
    &  -1.10 &   0.00 &   5 &  10 &  39 &    0.26 &    0.01 &    9.08 &    0.11  &  - & - & \textcolor{red}{-0.14} & 1 \\
    &  -1.10 &   0.00 &  10 &  15 &  34 &    0.25 &    0.01 &   11.85 &    0.20 &  \textcolor{red}{ -0.01} & 5 & \textcolor{red}{-0.16} & 1 \\
    &  -1.10 &   0.00 &  15 & 30 &  16 &    0.12 &    0.01 &   17.87 &    0.56 & \textcolor{red}{  -0.14} & 1 & \textcolor{red}{-0.23} & 1 \\
\hline
\hline
 Mg &  -2.50 &  -1.80 &   5 &  10 &  21 &    0.14 &    0.03 &    8.65 &    0.15  &  - & - & - & - \\
    &  -2.50 &  -1.80 &  10 &  15 &  46 &    0.24 &    0.02 &   11.47 &    0.15 &  0.1 & 15 &  - & -  \\
    &  -2.50 &  -1.80 &  15 & 30 &  22 &    0.15 &    0.03 &   18.51 &    0.49 &    0.01 & $>20$ & - & - \\
    &  -1.80 &  -1.10 &   5 &  10 &  50 &    0.12 &    0.01 &    9.01 &    0.09  &  - & - & -0.02 & $> 20$ \\
    &  -1.80 &  -1.10 &  10 &  15 & 108 &    0.09 &    0.01 &   11.96 &    0.10 & \textcolor{red}{  -0.03} & 10 & \textcolor{red}{-0.16} & 1 \\
    &  -1.80 &  -1.10 &  15 & 30 &  59 &    0.06 &    0.01 &   17.38 &    0.20 & \textcolor{red}{  -0.06} & 1 & \textcolor{red}{-0.08} & 1 \\
    &  -1.10 &   0.00 &   5 &  10 &  39 &    0.18 &    0.01 &    9.08 &    0.11  &  - & - & \textcolor{red}{0.04} & 1 \\
    &  -1.10 &   0.00 &  10 &  15 &  34 &    0.08 &    0.02 &   11.85 &    0.20 & -0.1 & 20 & \textcolor{red}{-0.16} & 1  \\
    &  -1.10 &   0.00 &  15 & 30 &  16 &   -0.02 &    0.02 &   17.87 &    0.56 & \textcolor{red}{  -0.20} & 1 & \textcolor{red}{-0.16} & 1 \\
\hline
\hline
 Ca &  -2.50 &  -1.80 &   5 &  10 &  21 &    0.27 &    0.03 &    8.65 &    0.15  &  -  & - & - & - \\
    &  -2.50 &  -1.80 &  10 &  15 &  46 &    0.20 &    0.03 &   11.47 &    0.15 & \textcolor{red}{-0.07} & 1 & - & -  \\
    &  -2.50 &  -1.80 &  15 & 30 &  22 &    0.28 &    0.03 &   18.51 &    0.42 &   -0.01 & $> 20$ & - & - \\
    &  -1.80 &  -1.10 &   5 &  10 &  50 &    0.16 &    0.01 &    9.07 &    0.09  &  - & - & \textcolor{red}{-0.11} & 1 \\
    &  -1.80 &  -1.10 &  10 &  15 & 105 &    0.15 &    0.01 &   11.96 &    0.10 &   -0.01 & $> 20$ & \textcolor{red}{-0.05} & 1  \\
    &  -1.80 &  -1.10 &  15 & 30 &  58 &    0.11 &    0.01 &   17.67 &    0.22 & \textcolor{red}{ -0.05} & 5 & \textcolor{red}{-0.16} & 1 \\
    &  -1.10 &   0.00 &   5 &  10 &  39 &    0.12 &    0.01 &    9.08 &    0.11  &  - & - & \textcolor{red}{-0.15} & 1 \\
    &  -1.10 &   0.00 &  10 &  15 &  35 &    0.08 &    0.01 &   11.84 &    0.20 & -0.04 & $> 20$ & \textcolor{red}{-0.12} & 1 \\
    &  -1.10 &   0.00 &  15 & 30 &  16 &    0.03 &    0.01 &   17.87 &    0.56 & \textcolor{red}{  -0.08} & 1 & \textcolor{red}{-0.24} & 1  \\
\hline
\hline
  S &  -2.50 &  -1.80 &   5 &  10 &  22 &    0.57 &    0.02 &    8.69 &    0.15  &  - & - & - & - \\
    &  -2.50 &  -1.80 &  10 &  15 &  46 &    0.53 &    0.02 &   11.53 &    0.16 &   \textcolor{red}{-0.04} & 10 & - & - \\
    &  -2.50 &  -1.80 &  15 & 30 &  22 &    0.57 &    0.03 &   18.51 &    0.42 &    0.00 & $> 20$ & - & - \\
    &  -1.80 &  -1.10 &   5 &  10 &  51 &    0.44 &    0.01 &    9.07 &    0.09  &  - & - & \textcolor{red}{-0.13} & 1 \\
    &  -1.80 &  -1.10 &  10 &  15 & 107 &    0.40 &    0.01 &   12.01 &    0.10 &  -0.04 & $> 20$ & \textcolor{red}{-0.14} & 1 \\
    &  -1.80 &  -1.10 &  15 & 30 &  62 &    0.36 &    0.01 &   17.75 &    0.22 & \textcolor{red}{  -0.08} & 1 & \textcolor{red}{-0.21} & 1  \\
    &  -1.10 &   0.00 &   5 &  10 &  39 &    0.36 &    0.01 &    9.08 &    0.11  &  - & - & \textcolor{red}{-0.21} & 1 \\
    &  -1.10 &   0.00 &  10 &  15 &  35 &    0.33 &    0.01 &   11.84 &    0.20 &   -0.03 & 15 & \textcolor{red}{-0.21} & 1  \\
    &  -1.10 &   0.00 &  15 & 30 &  16 &    0.19 &    0.02 &   17.87 &    0.56 & \textcolor{red}{  -0.17} & 5 & \textcolor{red}{-0.38} & 1 \\
\hline
\hline
 Si &  -2.50 &  -1.80 &   5 &  10 &  22 &    0.35 &    0.01 &    8.69 &    0.15  &  - & - & - & - \\
    &  -2.50 &  -1.80 &  10 &  15 &  47 &    0.37 &    0.01 &   11.47 &    0.15 &    0.02 & $> 20$ & - & -  \\
    &  -2.50 &  -1.80 &  15 & 30 &  23 &    0.30 &    0.01 &   18.42 &    0.39 &   \textcolor{red}{-0.05} & 10 & - & - \\
    &  -1.80 &  -1.10 &   5 &  10 &  52 &    0.34 &    0.01 &    9.07 &    0.09  &  - & - & 0.00 & $> 20$ \\
    &  -1.80 &  -1.10 &  10 &  15 & 109 &    0.32 &    0.00 &   11.96 &    0.10 &   -0.02 & $> 20$ & \textcolor{red}{-0.05} & 1 \\
    &  -1.80 &  -1.10 &  15 & 30 &  62 &    0.30 &    0.01 &   17.75 &    0.22 & \textcolor{red}{  -0.04} & 1 & 0.00 & $> 20$ \\
    &  -1.10 &   0.00 &   5 &  10 &  39 &    0.29 &    0.01 &    9.08 &    0.11  &  - & - & \textcolor{red}{-0.06} & 1 \\
    &  -1.10 &   0.00 &  10 &  15 &  35 &    0.26 &    0.01 &   11.84 &    0.20 &    -0.03 & $> 20$ & \textcolor{red}{-0.11} & 1 \\
    &  -1.10 &   0.00 &  15 & 30 &  16 &    0.22 &    0.02 &   17.87 &    0.56 & \textcolor{red}{  -0.07} & 10 & \textcolor{red}{-0.08} & 5 \\
\hline
\hline
 Ti &  -2.50 &  -1.80 &   5 &  10 &  18 &    0.10 &    0.04 &    8.65 &    0.16  &  - & - & - & - \\
    &  -2.50 &  -1.80 &  10 &  15 &  41 &    0.03 &    0.02 &   11.37 &    0.15 &   -0.07 & $> 20$ & - & -  \\
    &  -2.50 &  -1.80 &  15 & 30 &  21 &   -0.06 &    0.03 &   18.42 &    0.49 &  \textcolor{red}{-0.14} & 10 & - & -  \\
    &  -1.80 &  -1.10 &   5 &  10 &  49 &   -0.02 &    0.01 &    9.07 &    0.09  &  - & - & \textcolor{red}{-0.12} & 1 \\
    &  -1.80 &  -1.10 &  10 &  15 & 105 &   -0.07 &    0.01 &   11.96 &    0.10 & \textcolor{red}{  -0.05} & 1 & \textcolor{red}{-0.09} & 1  \\
    &  -1.80 &  -1.10 &  15 & 30 &  60 &   -0.10 &    0.01 &   17.67 &    0.22 & \textcolor{red}{  -0.08} & 1 & \textcolor{red}{-0.04} & 1 \\
    &  -1.10 &   0.00 &   5 &  10 &  39 &   -0.04 &    0.02 &    9.08 &    0.11  &  -  & - & \textcolor{red}{-0.14} & 5 \\
    &  -1.10 &   0.00 &  10 &  15 &  34 &   -0.05 &    0.01 &   11.85 &    0.20 &  -0.01 & 5 & \textcolor{red}{-0.07} & 1 \\ 
    &  -1.10 &   0.00 &  15 & 30 &  15 &   -0.12 &    0.02 &   17.59 &    0.55 & \textcolor{red}{  -0.08} & 5 & \textcolor{red}{-0.06} & 1  \\
\hline
\hline
\end{tabular}
\end{table*}

\begin{table*}
\contcaption{Median [X/Fe] and $r$ with their corresponding median absolute
deviation (MAD), evaluated in the three [M/H] and $r$ bins, and the
difference between each median and that corresponding to the lowest $r$
bin over each [M/H] range.
The significant differences indicated by the Kolmogorov-Smirnov test are in red, followed by the level of significance.}
\begin{tabular}{ccccccccccccccc}
\hline
elem & $\rm[M/H]_{l}$ & $\rm[M/H]_{u}$ & $r_{\rm l}$ & $r_{\rm u}$ & N & $<\rm[X/Fe]>$ & $e_{<\rm[X/Fe]>}$ & $<r>$ & $e_{<r>}$ & $\Delta <\rm[X/Fe]>_{[M/H]}$ & los (\%) & $\Delta <\rm[X/Fe]>_{r}$ & los (\%) \\
\hline
\hline
 Na &  -2.50 &  -1.80 &   5 &  10 &  14 &    1.01 &    0.08 &    8.69 &    0.19  &  - & - & - & - \\
    &  -2.50 &  -1.80 &  10 &  15 &  38 &    0.32 &    0.12 &   11.53 &    0.18 & \textcolor{red}{  -0.69} & 1 & - & - \\
    &  -2.50 &  -1.80 &  15 & 30 &  13 &    0.52 &    0.11 &   18.51 &    0.38 & \textcolor{red}{  -0.49} & 5 & - & - \\
    &  -1.80 &  -1.10 &   5 &  10 &  45 &   -0.07 &    0.07 &    9.01 &    0.11  &  - & - & \textcolor{red}{-1.08} & 1 \\
    &  -1.80 &  -1.10 &  10 &  15 & 89 &   -0.06 &    0.05 &   12.01 &    0.11 &  -0.01 & $> 20$ & \textcolor{red}{-0.37} & 1 \\
    &  -1.80 &  -1.10 &  15 & 30 &  48 &   -0.42 &    0.05 &   17.67 &    0.26 &  \textcolor{red}{-0.35} & 10 & \textcolor{red}{-0.93} & 1  \\
    &  -1.10 &   0.00 &   5 &  10 &  31 &   -0.03 &    0.05 &    9.00 &    0.14  &  - & - & \textcolor{red}{-1.04} & 1 \\
    &  -1.10 &   0.00 &  10 &  15 &  28 &   -0.31 &    0.02 &   12.56 &    0.22 & \textcolor{red}{  -0.28} & 1 & \textcolor{red}{-0.62} & 1 \\
    &  -1.10 &   0.00 &  15 & 30 &  13 &   -0.43 &    0.09 &   16.31 &    0.35 &  \textcolor{red}{-0.40} & 5 & \textcolor{red}{-0.94} & 1 \\
\hline
\hline
  N &  -2.50 &  -1.80 &   5 &  10 &  23 &    0.48 &    0.04 &    8.69 &    0.15  &  - & - & - & - \\
    &  -2.50 &  -1.80 &  10 &  15 &  47 &    0.54 &    0.02 &   11.47 &    0.15 &   0.06 & $> 20$ & - & - \\
    &  -2.50 &  -1.80 &  15 & 30 &  23 &    0.49 &    0.04 &   18.42 &    0.39 &    0.01 & $> 20$ & - & - \\
    &  -1.80 &  -1.10 &   5 &  10 &  53 &    0.13 &    0.02 &    9.07 &    0.09  &  - & - & \textcolor{red}{-0.34} & 1 \\
    &  -1.80 &  -1.10 &  10 &  15 & 110 &    0.19 &    0.01 &   12.01 &    0.10 & 0.06 & 15 & \textcolor{red}{-0.35} & 1  \\
    &  -1.80 &  -1.10 &  15 & 30 &  62 &    0.20 &    0.01 &   17.75 &    0.22 & \textcolor{red}{   0.07} & 5 & \textcolor{red}{-0.29} & 1 \\
    &  -1.10 &   0.00 &   5 &  10 &  40 &    0.02 &    0.01 &    9.11 &    0.11  &  - & - & \textcolor{red}{-0.46} & 1 \\
    &  -1.10 &   0.00 &  10 &  15 &  35 &    0.03 &    0.01 &   11.84 &    0.20 &    0.01 & $> 20$ & \textcolor{red}{-0.51} & 1 \\
    &  -1.10 &   0.00 &  15 & 30 &  16 &    0.06 &    0.03 &   17.87 &    0.56 &    0.04 & $> 20$ & \textcolor{red}{-0.43} & 1 \\
\hline
\hline
 Al &  -2.50 &  -1.80 &   5 &  10 &  20 &   -0.15 &    0.03 &    8.65 &    0.16  &  - & - & - & - \\
    &  -2.50 &  -1.80 &  10 &  15 &  45 &   -0.07 &    0.03 &   11.45 &    0.15 & \textcolor{red}{   0.08} & 1 & - & -  \\
    &  -2.50 &  -1.80 &  15 & 30 &  19 &   -0.13 &    0.03 &   18.51 &    0.45 &   0.02  & 15 & - & - \\
    &  -1.80 &  -1.10 &   5 &  10 &  47 &   -0.10 &    0.02 &    9.01 &    0.09  &  - & - & \textcolor{red}{0.05} & 5 \\
    &  -1.80 &  -1.10 &  10 &  15 & 101 &   -0.14 &    0.01 &   12.01 &    0.11 & \textcolor{red}{  -0.04} & 5 & \textcolor{red}{-0.08} & 1 \\
    &  -1.80 &  -1.10 &  15 & 30 &  56 &   -0.12 &    0.01 &   17.38 &    0.23 &  -0.02 & $> 20$ & \textcolor{red}{0.02} & 1 \\
    &  -1.10 &   0.00 &   5 &  10 &  39 &    0.10 &    0.03 &    9.08 &    0.11  &  - & - & \textcolor{red}{0.25} & 1 \\
    &  -1.10 &   0.00 &  10 &  15 &  34 &   0.01 &    0.04 &   11.84 &    0.20 & \textcolor{red}{  -0.09} & 5 & \textcolor{red}{0.08} & 5  \\
    &  -1.10 &   0.00 &  15 & 30 &  15 &   -0.12 &    0.04 &   17.59 &    0.55 & \textcolor{red}{  -0.22} & 10 & \textcolor{red}{0.01} & 5 \\
\hline
\hline
  C &  -2.50 &  -1.80 &   5 &  10 &  21 &    0.13 &    0.05 &    8.65 &    0.14  &  - & - & - & - \\
    &  -2.50 &  -1.80 &  10 &  15 &  46 &    0.11 &    0.04 &   11.53 &    0.16 &  \textcolor{red}{ -0.02} & 5 & - & - \\
    &  -2.50 &  -1.80 &  15 & 30 &  23 &    0.11 &    0.06 &   18.42 &    0.39 &  -0.02 & $> 20$ & - & - \\
    &  -1.80 &  -1.10 &   5 &  10 &  52 &   -0.20 &    0.02 &    9.07 &    0.09  &  - & - & \textcolor{red}{-0.33} & 1 \\
    &  -1.80 &  -1.10 &  10 &  15 & 109 &   -0.26 &    0.01 &   11.96 &    0.10 & \textcolor{red}{  -0.06} & 5 & \textcolor{red}{-0.37} & 1 \\
    &  -1.80 &  -1.10 &  15 & 30 &  62 &   -0.28 &    0.01 &   17.75 &    0.22 & \textcolor{red}{  -0.08} & 1 & \textcolor{red}{-0.40} & 1 \\
    &  -1.10 &   0.00 &   5 &  10 &  39 &    0.03 &    0.01 &    9.08 &    0.11  &  - & - & \textcolor{red}{-0.10} & 1 \\
    &  -1.10 &   0.00 &  10 &  15 &  35 &   -0.12 &    0.03 &   11.84 &    0.20 & \textcolor{red}{  -0.15} & 1 & \textcolor{red}{-0.24} & 1 \\
    &  -1.10 &   0.00 &  15 & 30 &  16 &   -0.26 &    0.04 &   17.87 &    0.56 & \textcolor{red}{  -0.29} & 1 & \textcolor{red}{-0.38} & 1 \\
\hline
\hline
  K &  -2.50 &  -1.80 &   5 &  10 &  17 &    0.12 &    0.04 &    8.65 &    0.16  &  - & - & - & - \\
    &  -2.50 &  -1.80 &  10 &  15 &  42 &    0.15 &    0.03 &   11.53 &    0.16 &    0.03 & $> 20$ & - & -  \\
    &  -2.50 &  -1.80 &  15 & 30 &  18 &    0.18 &    0.04 &   18.54 &    0.55 &    0.06 & $> 20$ & - & - \\
    &  -1.80 &  -1.10 &   5 &  10 &  51 &    0.03 &    0.03 &    9.07 &    0.09  &  - & - & -0.09 & 20 \\
    &  -1.80 &  -1.10 &  10 &  15 & 106 &   0.00 &    0.01 &   11.96 &    0.10 &   -0.03 & $> 20$ & \textcolor{red}{-0.15} & 1 \\
    &  -1.80 &  -1.10 &  15 & 30 &  56 &   0.00 &    0.02 &   17.67 &    0.22 &  -0.03 & 20 & \textcolor{red}{-0.18} & 1 \\
    &  -1.10 &   0.00 &   5 &  10 &  38 &    0.02 &    0.02 &    9.11 &    0.11  &  - & - & \textcolor{red}{-0.11} & 1 \\
    &  -1.10 &   0.00 &  10 &  15 &  35 &   -0.07 &    0.01 &   11.84 &    0.20 &  -0.09 & $> 20$ & \textcolor{red}{-0.22} & 1 \\
    &  -1.10 &   0.00 &  15 & 30 &  15 &   -0.12 &    0.03 &   17.87 &    0.58 & \textcolor{red}{  -0.14} & 5 & \textcolor{red}{-0.30} & 1 \\
\hline
\hline
 Mn &  -2.50 &  -1.80 &   5 &  10 &  20 &   -0.09 &    0.04 &    8.69 &    0.16  &  - & - & - & - \\
    &  -2.50 &  -1.80 &  10 &  15 &  45 &    0.03 &    0.02 &   11.45 &    0.15 & \textcolor{red}{   0.13} & 1 & - & - \\
    &  -2.50 &  -1.80 &  15 & 30 &  23 &   -0.08 &    0.02 &   18.42 &    0.39 &    0.021 & $> 20$ & - & - \\
    &  -1.80 &  -1.10 &   5 &  10 &  50 &   -0.18 &    0.01 &    9.10 &    0.09  &  - & - & \textcolor{red}{-0.09} & 1 \\
    &  -1.80 &  -1.10 &  10 &  15 & 105 &   -0.21 &    0.01 &   12.02 &    0.11 &   \textcolor{red}{0.01} & 5 & \textcolor{red}{-0.24} & 1 \\
    &  -1.80 &  -1.10 &  15 & 30 &  61 &   -0.24 &    0.01 &   17.75 &    0.22 &  \textcolor{red}{ -0.01} & 1 & \textcolor{red}{-0.16} & 1 \\
    &  -1.10 &   0.00 &   5 &  10 &  38 &   -0.14 &    0.01 &    9.11 &    0.11  &  - & - & \textcolor{red}{-0.05} & 1 \\
    &  -1.10 &   0.00 &  10 &  15 &  33 &   -0.17 &    0.01 &   11.48 &    0.19 &   -0.01 & $> 20$ & \textcolor{red}{-0.20} & 1 \\
    &  -1.10 &   0.00 &  15 & 30 &  15 &   -0.18 &    0.02 &   17.59 &    0.55 &  -0.02 & $> 20$ & \textcolor{red}{-0.10} & 10 \\
\hline
\hline
\end{tabular}
\end{table*}

\begin{table*}
\contcaption{Median [X/Fe] and $r$ with their corresponding median absolute
deviation (MAD), evaluated in the three [M/H] and $r$ bins, and the
difference between each median and that corresponding to the lowest $r$
bin over each [M/H] range. The significant differences indicated by the
Kolmogorov-Smirnov test are in red, followed by the level of
significance.}
\begin{tabular}{cccccccccccccc}
\hline
elem & $\rm[M/H]_{l}$ & $\rm[M/H]_{u}$ & $r_{\rm l}$ & $r_{\rm u}$ & N & $<\rm[X/Fe]>$ & $e_{<\rm[X/Fe]>}$ & $<r>$ & $e_{<r>}$ & $\Delta <\rm[X/Fe]>_{[M/H]}$ & los (\%) & $\Delta <\rm[X/Fe]>_{r}$ & los (\%) \\
\hline
\hline
 Ni &  -2.50 &  -1.80 &   5 &  10 &  20 &   -0.07 &    0.02 &    8.65 &    0.15  &  - & - & - & - \\
    &  -2.50 &  -1.80 &  10 &  15 &  46 &   -0.05 &    0.01 &   11.47 &    0.15 &  0.02 &  $> 20$ & - & - \\
    &  -2.50 &  -1.80 &  15 & 30 &  22 &   -0.05 &    0.02 &   18.51 &    0.49 &    0.02 & $> 20$ & - & - \\
    &  -1.80 &  -1.10 &   5 &  10 &  51 &   -0.07 &    0.01 &    9.07 &    0.09  &  - & - & \textcolor{red}{-0.01} & 1 \\
    &  -1.80 &  -1.10 &  10 &  15 & 108 &   -0.07 &    0.00 &   12.01 &    0.10 &    0.00 & $> 20$ & \textcolor{red}{-0.02} & 1 \\
    &  -1.80 &  -1.10 &  15 & 30 &  58 &   -0.08 &    0.01 &   17.38 &    0.21 &    -0.01 & $> 20$ & \textcolor{red}{-0.04} & 5 \\
    &  -1.10 &   0.00 &   5 &  10 &  39 &   -0.05 &    0.01 &    9.08 &    0.11  &  - & - & 0.02 & $> 20$ \\
    &  -1.10 &   0.00 &  10 &  15 &  34 &   -0.05 &    0.01 &   11.85 &    0.20 &   0.00 & $> 20$ & 0.00 & $> 20$ \\
    &  -1.10 &   0.00 &  15 & 30 &  15 &   -0.13 &    0.02 &   17.59 &    0.55 & \textcolor{red}{  -0.08} & 5 & -0.08 & $> 20$ \\
\hline
\hline
  V &  -2.50 &  -1.80 &   5 &  10 &  5 &   -0.55 &    0.05 &    7.99 &    0.20  &  - & - & - & - \\
    &  -2.50 &  -1.80 &  10 &  15 &  22 &   -0.32 &    0.05 &   11.68 &    0.26 &  \textcolor{red}{  0.23} & 1 & - & - \\
    &  -2.50 &  -1.80 &  15 & 30 &  11 &   -0.12 &    0.09 &   18.61 &    0.73 & \textcolor{red}{   0.43} & 1 & - & - \\
    &  -1.80 &  -1.10 &   5 &  10 &  41 &   -0.49 &    0.03 &    9.07 &    0.08  &  - & - & \textcolor{red}{0.07} & 1 \\
    &  -1.80 &  -1.10 &  10 &  15 & 89 &   -0.46 &    0.02 &   12.06 &    0.11 & 0.03 & $> 20$ & \textcolor{red}{-0.14} & 1 \\
    &  -1.80 &  -1.10 &  15 & 30 &  52 &   -0.40 &    0.03 &   17.67 &    0.23 & \textcolor{red}{   0.09} & 10 & \textcolor{red}{-0.29} & 1 \\
    &  -1.10 &   0.00 &   5 &  10 &  33 &   -0.33 &    0.05 &    8.83 &    0.14  &  - & - & \textcolor{red}{0.22} & 1 \\
    &  -1.10 &   0.00 &  10 &  15 &  30 &   -0.46 &    0.05 &   12.60 &    0.23 &  -0.13 & $> 20$ & \textcolor{red}{-0.14} & 1 \\
    &  -1.10 &   0.00 &  15 & 30 &  15 &   -0.41 &    0.04 &   17.59 &    0.55 &    -0.08 & $> 20$ & \textcolor{red}{-0.29} & 1 \\
\hline
\hline
\end{tabular}
\end{table*}

%%%%%%%%%%%%%%%%%%%%%%%%%%%%%%%%%%%%%%%%%%%%%%%%%%

% Don't change these lines
\bsp	% typesetting comment
\label{lastpage}
\end{document}